\documentclass[useAMS]{mn2e}
\usepackage{graphicx}
\usepackage{color}
\usepackage{amssymb}
\title[Lithium abundance in F and G supergiants]
{LITHIUM ABUNDANCE IN ATMOSPHERES OF F- AND G-TYPE SUPERGIANTS AND BRIGHT GIANTS}

\author[L.S. Lyubimkov et al.]
{Leonid~S.~Lyubimkov,$^1$\thanks{E-mail: lyub@crao.crimea.ua (LSL); dll@astro.as.utexas.edu (DLL);
bogdan@mao.kiev.ua (BMK); yp@mao.kiev.ua (YVP)} David~L.~Lambert,$^2$$^\star$ Bogdan M. Kaminsky,$^3$$^\star$ \and Yakov V. Pavlenko,$^3$$^\star$ Dmitry~B.~Poklad,$^1$ Tamara~M.~Rachkovskaya$^1$\\
$^1$Crimean Astrophysical Observatory, Ukraine\\
$^2$The W.J. McDonald Observatory, The University of Texas at Austin, USA\\
$^{3}$Main Astronomical Observatory of the Academy of Sciences of Ukraine\\} 

\begin{document}

\date{Accepted. Received ; in original form}

\pagerange{\pageref{firstpage}--\pageref{lastpage}} \pubyear{}

\maketitle

\label{firstpage}

\begin{abstract}
Lithium in the atmosphere of a F or G supergiant reflects the initial Li abundance and the internal history of the star. During evolution of a star from the main sequence (MS) to the supergiant phase, lithium may be destroyed by, for example, rotationally-induced mixing in the MS stars and strongly diluted by development of the supergiant's convective envelope. In order to probe the connection between atmospheric Li abundance and evolutionary predictions, we present a non-LTE abundance analysis of the resonance doublet Li I at 6707.8 \AA\ for 55 Galactic F and G supergiants and bright giants (we observed 43 of them, the remaining 12 are added from Luck and Wepfer's list). The derived lithium abundances $\log\epsilon$(Li) may be considered in three groups, namely: (i) ten Li-rich giants with $\log\epsilon$(Li) = 2.0 to 3.2 (all ten are F-type or A9 stars); (ii) thirteen G- to K0-type stars with Li abundances in the narrow range $\log\epsilon$(Li) = 1.1-1.8; (iii) all other stars provide just upper limits to the Li abundance.

The derived Li abundances are compared with theoretical predictions of 2 to 15 M$_\odot$ stars (both rotating and non-rotating). Our results are generally in good agreement with theory. In particular, the absence of detectable lithium for the majority of programme stars is explainable. The comparison suggests that the stars may be separated by mass $M$ into two groups, namely $M \lesssim $~6~M$_\odot$ and $M >$~6~M$_\odot$. All Li-rich giants and supergiants with $\log\epsilon$(Li) $\ge$ 2.0 have masses $M <$~6~M$_\odot$; this conclusion follows not only from our work but also from a scrutiny of published data. Eleven of thirteen stars with $\log\epsilon$(Li) = 1.1-1.8, specifically the stars with $M <$~6~M$_\odot$, show good agreement with the post-first dredge-up surface abundance $\log\epsilon$(Li) $\approx$ 1.4 predicted for the non-rotating 2-6 M$_\odot$ stellar models. An absence of Li-rich stars in the range $M >$~6~M$_\odot$ agrees with the theoretical prediction that F and G supergiants and giants with $M >$~6~M$_\odot$ cannot show detectable lithium.
     
We note that present theory appears unable to account for the derived Li abundances for some stars, namely for (i) a few relatively low-mass Li-rich giants ($M <$~6~M$_\odot$), whose high Li abundances accompanied by rather high rotational velocities or substantial nitrogen excess contradict theoretical predictions; (ii) the relatively high-mass supergiants HR 461 and HR 8313 ($M >$~6~M$_\odot$) with the detected abundances $\log\epsilon$ = 1.3-1.5. It is possible that the lithium in such stars was synthesized recently.

\end{abstract}


% #1
\section{Introduction}

     Lithium is one of a few key chemical elements, whose observed atmospheric abundance reflects the evolution of a star. The Li abundance in young stars of normal (solar) metallicity is $\log\epsilon$(Li) = 3.2$\pm$0.1 (see, e.g., Balachandran et al. 2011, Mallik et al. 2010 and Randich 2010). This value seems to be in excellent agreement with the meteoritic abundance $\log\epsilon$(Li) = 3.26$\pm$0.05 (Asplund et al. 2009). One supposes that the value $\log\epsilon$(Li) = 3.2$\pm$0.1 is the lithium content in stars of the solar vicinity at the beginning of their evolution. Note that the Li abundance $\log\epsilon$(Li) here and subsequently is given on the standard logarithmic scale, where the hydrogen abundance is $\log\epsilon$(H) = 12.00. 

      The lithium abundance is known mainly for F- and later type stars, where Li~I lines (mostly the resonance doublet at 6708~\AA) can be observed. (Sometimes the Li~I 6708 line is observed in spectra of magnetic Ap-stars with an enhanced Li content.) For F, G and K dwarfs a correlation exists between $\log\epsilon$(Li) and the effective temperature $T_{\rm eff}$: the Li abundance tends to decrease with decreasing $T_{\rm eff}$. In particular, lithium depletion is reliably established for the Sun, namely $\log\epsilon_\odot$(Li) = 1.05$\pm$0.10 according to Asplund et al. (2009) and 1.03$\pm$0.03 according to Caffau et al. (2010), but the meteoritic abundance is 3.26, so there is a Li depletion by $\sim$ 2 dex. The lower $\log\epsilon$(Li) in atmospheres of the Sun and other cool dwarfs is attributed principally to lithium destruction during the main sequence (MS) evolutionary phase. Surface Li depletion for stars with masses $M <$ 1.2 M$_\odot$ can start in the pre-MS phase (see, e.g., Maeder 2009, Sec. 20.7). 

      The observed Li abundances in atmospheres of F- and G-type supergiants and giants, the stars of interest here, show a large scatter from star to star. Most stars display a significantly lower Li content as compared with young stars; moreover, the Li~I 6708 line is often not seen, so only an upper limit to $\log\epsilon$(Li) can be evaluated (see, e.g., the early study of Luck 1977). On the other hand, there are a few F, G and K supergiants and giants with rather high lithium abundances $\log\epsilon$(Li) $\approx$ 2-4. Following some authors, we shall use the name 'Li-rich' for stars with $\log\epsilon$(Li) $\ge$ 2.0. 

      As is well known, the accuracy of two fundamental stellar parameters, namely the effective temperature $T_{\rm eff}$ and surface gravity $\log g$, plays an important role in the precision of an abundance analysis. Recently, we determined accurate fundamental parameters for 63 Galactic A-, F- and G-type supergiants and bright giants (i.e., luminosity classes I and II) of masses about three solar masses or greater, including $T_{\rm eff}$, $\log g$ and the iron abundance $\log\epsilon$(Fe) (Lyubimkov et al. 2010, hereinafter Paper I). A significant improvement in the accuracy of the $\log g$ values was obtained through application of van Leeuwen's (2007) new reduction of the Hipparcos parallaxes (see Lyubimkov et al. 2009 and Paper I for details). We found that the typical error in our $\log g$ values is $\pm$0.06 dex for stars with distances $d <$ 300 pc and $\pm$0.12 dex for stars with $d$ between 300 and 700 pc. 
For more distant supergiants with $d >$ 700 pc, where parallaxes are uncertain or unknown, the typical errors in $\log g$ are 0.2-0.3 dex. 
In order to check the derived $T_{\rm eff}$ values, a comparison with results inferred with the infrared flux method was used; the mean error of $\pm$120~K in $T_{\rm eff}$ for stars with $d <$ 700 pc was found. 

      Using these new stellar parameters, we determined the nitrogen abundance in the atmospheres of 30 Galactic A and F supergiants (Lyubimkov et al. 2011). We confirmed the surface N enrichment previously reported for such supergiants and showed that rotationally-induced mixing during the main sequence (MS) evolutionary phase can play an important role in this enrichment. The mixing leads to the transfer of the CNO-cycled material, i.e., He, C, N and O, from the stellar interior to the surface. Apart from the observed N enrichment and associated C deficiency in atmospheres of A, F and G supergiants, a lithium deficiency may confirm the mixing. According to theory, Li depletion (also, Be and B depletion) due to the mixing in the MS phase can be very strong (Heger \& Langer 2000; Frischknecht et al. 2010). Moreover, one anticipates that significant Li depletions may occur, even when marked changes in the surface N and C abundances are absent (e.g., for stars with relatively low masses and/or low rotational velocities). Therefore, Li is a more sensitive indicator of the mixing than elements of the CNO-cycle. 

      Thus, using the new parameters $T_{\rm eff}$ and $\log g$, we analyzed the Li abundance for 43 F- and G-type supergiants and bright giants. In Section 2 we describe observations and selection of programme stars. In Section 3 we discuss computations of Li I lines, as well as synthetic spectra in the region of the Li I 6708 \AA\ resonance doublet, which is a principal source of the Li abundance derivation. Calculations of the model atmospheres are described there, too. In Section 4 we present in detail the Li abundance determination, which is based on a comparison of observed and synthetic spectra in the region of the Li I 6708 \AA\ line. Both LTE and non-LTE analyses of this line are implemented (LTE is local thermodynamic equilibrium). In Section 5 we compare the derived Li abundances with other published data. Special attention is given to three Li-rich stars, namely HD 17905, HR 7008 and HR 3102. In Section 5 additional twelve bright giants from Luck \& Wepfer (1995) list are selected. Their parameters $T_{\rm eff}$ and $\log g$ are redetermined and their Li abundances are corrected. In Section 7 the Li abundance is analyzed as a function of effective temperature $T_{\rm eff}$. In Section 8 the Li abundance is considered as a function of another basic parameter, the stellar mass $M$. The masses of Li-rich stars are of special interest there. In Section 9 we compare the derived Li abundances with recent theoretical predictions. We show that both the predictions and observations reveal an evident difference between stars with masses $M \lesssim $~6~M$_\odot$ and $M~>$~6~M$_\odot$. Concluding remarks are presented in Section 10.

% Table-1
\begin{table*}
 \centering
 \begin{minipage}{160mm}
 \caption{Basic parameters and the Li abundance for 43 programme stars (first part of the table contains 
the stars with the detectable Li abundances, second one shows the stars with the Li upper limits)}
  \begin{tabular}{|c|c|c|c|c|c|c|c|c|c|c|c|}
  \hline
HR  & HD  & Sp & $T_{\rm eff}$ & $\log g$ & $V_{t}$, & $d$, pc & $M/M_\odot$ & $\log\epsilon$(Fe) & $v \sin i$, & $\log\epsilon$(Li) & $\log\epsilon$(Li)   \\
    &     &    &               &          & km~s$^{-1}$&       &             &                    & km~s$^{-1~a)}$ &    (LTE)                            &     (non-LTE)        \\
  \hline
461 & 9900  &  K0 Ia     &     4430     &   1.18    &   2.8    &   529   &    9.5     &     7.47     &9&      0.89       &      1.27$\pm$0.19  \\
792 & 16780 &  G5 II     &     5020     &   2.09    &   2.9    &   397   &    5.0     &     7.49     &10&      1.34       &      1.53$\pm$0.14  \\
 -- & 17905 &  F5 III    &     6580     &   3.26    &   2.5    &   157   &    2.4     &     7.55     &53&      3.14       &      3.18$\pm$0.11  \\
1327& 27022 &  G0 II     &     5440     &   2.89    &   1.2    &   98    &    2.8     &     7.41     &6&      1.4        &      1.51$\pm$0.21  \\
2833& 58526 &  G3 Ib     &     5380     &   2.21    &   4.0    &   375   &    4.7     &     7.58     &11&      1.16       &      1.28$\pm$0.16  \\
3102& 65228 &  F7 II     &     5690     &   2.17    &   3.7    &   161   &    5.1     &     7.61     &12&      2.27       &      2.33$\pm$0.19  \\
4786& 109379&  G5 II     &     5100     &   2.52    &   1.5    &   45    &    3.7     &     7.60     &8&      1.01       &      1.16$\pm$0.10  \\
7008& 172365&  F8 II     &     6220     &   2.53    &   3.1    &   342   &    4.2     &     7.60     &58&      3.1        &      3.07$\pm$0.12  \\
8313& 206859&  G5 Ib     &     4910     &   1.58    &   2.8    &   283   &    7.1     &     7.48     &10&      1.39       &      1.53$\pm$0.14  \\
8412& 209693&  G5 Ia     &     5280     &   2.35    &   2.3    &   284   &    4.2     &     7.55     &9&      1.34       &      1.47$\pm$0.16  \\
8692& 216206&  G4 Ib     &     4960     &   1.90    &   3.4    &   413   &    5.6     &     7.40     &9&      1.06       &      1.26$\pm$0.14  \\ 
  \hline
157 & 3421  &  G2.5 IIa  &     5130     &   2.15    &   2.5    &   259   &    4.8     &     7.41     &9&  $\le$ 0.9     &      $\le$ 1.0       \\
207 & 4362  &  G0 Ib     &     5220     &   1.55    &   4.0    &   935   &    7.9     &     7.38     &10.5&  $\le$ 0.5     &      $\le$ 0.6       \\ 
849 & 17818 &  G5 Iab:   &     5020     &   1.73    &   2.2    &   538   &    6.5     &     7.58     &10.5&  $\le$ 0.8     &      $\le$ 0.9       \\
1242& 25291 &  F0 II     &     6815     &   1.87    &   3.2    &   629   &    8.3     &     7.43     &7.5&  $\le$ 1.9     &      $\le$ 1.9       \\
1270& 25877 &  G8 IIa    &     5060     &   1.91    &   1.7    &   427   &    5.7     &     7.59     &8&  $\le$ 0.7     &      $\le$ 0.8       \\
1303& 26630 &  G0 Ib     &     5380     &   1.73    &   3.6    &   275   &    7.0     &     7.41     &10.5&  $\le$ 0.6     &      $\le$ 0.7       \\
1603& 31910 &  G1 Ib-IIa &     5300     &   1.79    &   4.8    &   265   &    6.5     &     7.46     &11&  $\le$ 0.8     &      $\le$ 0.9       \\
1829& 36079 &  G0 II     &     5450     &   2.60    &   1.3    &    49   &    3.5     &     7.41     &9&  $\le$ 0.7     &      $\le$ 0.8       \\
2000& 38713 &  G5 II     &     5000     &   2.45    &   2.1    &   224   &    3.9     &     7.40     &6.5&  $\le$ 0.8     &      $\le$ 1.0       \\ 
2453& 47731 &  G5 Ib     &     4900     &   1.70    &   2.3    &   633   &    6.5     &     7.62     &10.5&  $\le$ 0.7     &      $\le$ 0.9       \\
2693& 54605 &  F8 Ia     &     5850     &   1.00    &   7.0    &   495   &    14.9    &     7.51     &15&  $\le$ 1.5     &      $\le$ 1.5       \\
2786& 57146 &  G2 Ib     &     5260     &   1.90    &   3.2    &   386   &    5.9     &     7.55     &10.5&  $\le$ 1.0     &      $\le$ 1.1       \\
2881& 59890 &  G3 Ib     &     5300     &   1.66    &   5.2    &   461   &    7.3     &     7.33     &11&  $\le$ 0.8     &      $\le$ 0.9       \\
3045& 63700 &  G6 Iab-Ib &     4880     &   1.21    &   5.1    &   370   &    9.9     &     7.43     &10.5&  $\le$ 0.6     &      $\le$ 0.8       \\
3188& 67594 &  G2 Ib     &     5210     &   1.75    &   3.3    &   325   &    6.6     &     7.51     &10&  $\le$ 0.5     &      $\le$ 0.6       \\ 
3229& 68752 &  G5 II     &     5130     &   2.04    &   2.3    &   267   &    5.2     &     7.51     &10&  $\le$ 0.8     &      $\le$ 0.9       \\
3459& 74395 &  G1 Ib     &     5370     &   2.08    &   3.5    &   236   &    5.2     &     7.53     &10.5&  $\le$ 0.9     &      $\le$ 1.0       \\
4166& 92125 &  G0 II     &     5475     &   2.36    &   2.7    &   177   &    4.2     &     7.52     &9.5&  $\le$ 0.9     &      $\le$ 1.0       \\
5143& 119035&  G5 II:    &     5190     &   2.75    &   1.0    &   166   &    3.2     &     7.32     &6&  $\le$ 0.5     &      $\le$ 0.6       \\ 
5165& 119605&  G0 Ib-IIa &     5430     &   2.37    &   2.5    &   253   &    4.2     &     7.33     &10.5&  $\le$ 0.9     &      $\le$ 1.0       \\
6536& 159181&  G2 Ib-IIa &     5160     &   1.86    &   3.0    &   117   &    6.0     &     7.52     &11&  $\le$ 0.9     &      $\le$ 1.0       \\
6978& 171635&  F7 Ib     &     6000     &   1.70    &   4.6    &   649   &    8.2     &     7.41     &10&  $\le$ 1.5     &      $\le$ 1.5       \\
7164& 176123&  G3 II     &     5200     &   2.25    &   2.5    &   334   &    4.5     &     7.40     &7&  $\le$ 0.7     &      $\le$ 0.8       \\
7456& 185018&  G0 Ib     &     5550     &   2.06    &   2.8    &   370   &    5.5     &     7.34     &10&  $\le$ 0.9     &      $\le$ 1.0       \\
7542& 187203&  F8 Ib-II  &     5750     &   2.15    &   4.2    &   376   &    5.3     &     7.67     &25&  $\le$ 1.5     &      $\le$ 1.5       \\
7770& 193370&  F5 Ib     &     6180     &   1.53    &   5.0    &   960   &    10.0    &     7.28     &3.5&  $\le$ 1.6     &      $\le$ 1.6       \\
7795& 194069&  G9 Ib-II  &     4870     &   2.00    &   3.1    &   403   &    5.3     &     7.50     &9.5&  $\le$ 0.6     &      $\le$ 0.7       \\
7796& 194093&  F8 Ib     &     5790     &   1.02    &   5.2    &   562   &    14.5    &     7.46     &9&  $\le$ 1.5     &      $\le$ 1.5       \\
7834& 195295&  F5 II     &     6570     &   2.32    &   3.6    &   235   &    5.3     &     7.50     &8.5&  $\le$ 1.9     &      $\le$ 1.9       \\
7847& 195593&  F5 Iab    &     6290     &   1.44    &   4.1    &   1040  &    11.2    &     7.44     &6&  $\le$ 1.6     &      $\le$ 1.6       \\
8232& 204867&  G0 Ib     &     5490     &   1.86    &   3.7    &   165   &    6.4     &     7.60     &10.5&  $\le$ 1.0     &      $\le$ 1.1       \\
8414& 209750&  G2 Ib     &     5210     &   1.76    &   3.8    &   161   &    6.5     &     7.53     &10.5&  $\le$ 0.9     &      $\le$ 0.9       \\
   \hline
   \end{tabular}
	$^{a)}$ These $v \sin i$ values should be considered as the upper limits for the actual rotational velocities, because they are derived for $V_{mac} = 0$.
 \end{minipage}
\end{table*}

% #2
\section{OBSERVATIONS AND THE STELLAR SAMPLE}

      Programme stars are selected from Paper I. High-resolution spectra of the stars in Paper I were acquired at the McDonald observatory mainly in 2003--2004 using the cross-dispersed echelle coud\'e spectrograph at the 2.7-m telescope (Tull et al. 1995). Subsequently, we observed at the same telescope additionally two Li-rich F-type stars, namely HR 7008 and HD 17905; these observations were obtained in June and November 2009, respectively. All the stars were observed at a resolving power of $R$ = 60000; the signal-to-noise ratio was between 100 and 450. Reductions of the CCD spectral images were performed using standard IRAF routines.  

      The list of the stars selected from Paper I for lithium analysis is presented in Table~1. All the G-type stars from Paper I were included; the majority of the F-type stars with distances $d < 700$~pc and two F stars with $d > 700$~pc were included (see Tables~2 and 3 in Paper I). The programme stars in Table~1 are mostly F and G supergiants and bright giants (luminosity classes I and II), excluding the K0 Ia supergiant HR~461 and the F5~III giant HD~17905. Note that for HR~461 its earlier classification G5~II in the Bright Star Catalogue has been corrected in Paper I.  The first (upper) part of Table~1 contains the stars with a detectable Li~I 6708 line; its second (lower) part includes the stars, where this line is not seen. We show in Table~1 the HR and HD numbers, spectral type and luminosity class (Sp), effective temperature $T_{\rm eff}$, surface gravity $\log g$, microturbulent parameter $V_t$, distance $d$, iron abundance $\log \epsilon$(Fe) and mass $M$ in reference to the solar mass M$_\odot$. All the parameters are taken from Paper I, apart from two new objects, i.e. HR~7008 and HD~17905 (determination of their parameters is considered below). Errors in the parameters $T_{\rm eff}$, $\log g$, $\log \epsilon$(Fe) and $M$ for individual stars are presented in Paper I. Note that the mean error in the masses M for the stars in Table~1 is $\pm$8 per cent.

      We enlarged our sample adding 12 bright giants from Luck \& Wepfer's (1995) list. The star's parameters $T_{\rm eff}$ and $\log g$ were redetermined with our method; the Li abundances were corrected and, in particular, non-LTE corrections were included (see Section 6 for details).

% #3
\section{ COMPUTATIONS OF LITHIUM LINES, SYNTHETIC SPECTRA AND MODEL ATMOSPHERES}
% #3.1
\subsection{Spectral line data}

% Table-2
\begin{table}
  \caption{List of spectral lines in the vicinity of the Li~I 6708 resonance doublet}
  \centering
    \begin{tabular}{|l|c|c|c|}
    \hline
     Atom & $\lambda,$ \AA & $\log$ \textit{gf} & $E$, eV\\
     or molecule&&&\\
     \hline
    Fe I &6703.567&-3.090&2.758\\
    Fe I &6704.481&-2.660&4.217\\
    Fe I &6705.101&-1.155&4.607\\
    Si I &6706.980&-2.480&5.954\\
    CN   &6707.320&-1.849&1.206\\
    Sm II&6707.342&-2.000&0.884\\
    Er II&6707.418&-1.441&3.480\\
    Fe I &6707.432&-2.350&4.607\\
    Nd II&6707.433&-2.181&1.500\\
    Nd II&6707.453&-3.181&2.880\\
    Gd II&6707.462&-1.981&3.270\\
    Sm II&6707.473&-1.480&0.930\\
    V I  &6707.518&-1.995&2.743\\
    CN   &6707.529&-1.897&1.881\\
    CN   &6707.529&-1.897&1.894\\
    CN   &6707.529&-1.503&0.956\\
    Sm II&6707.648&-1.270&1.746\\
    CN   &6707.675&-3.205&0.956\\
    Sc I &6707.752&-2.672&4.049\\
    Li I &6707.754&-0.431&0.000\\
    Nd II&6707.755&-3.550&0.170\\
    Li I &6707.766&-0.209&0.000\\
    Sm II&6707.779&-2.680&2.037\\
    Li I &6707.904&-0.733&0.000\\
    Li I &6707.917&-0.510&0.000\\
    Nd II&6708.030&-1.130&1.520\\
    Ce II&6708.077&-2.570&2.250\\
    Er II&6708.088&-2.581&3.150\\
    V I  &6708.094&-3.113&1.218\\
    Ce II&6708.099&-2.120&0.700\\
    Nd II&6708.400&-2.481&3.190\\
    Nd II&6708.458&-1.081&3.530\\
    CN   &6708.500&-1.900&1.868\\
    CN   &6708.972&-1.711&0.886\\
    Fe I &6710.319&-4.880&1.485\\
    Fe I &6713.046&-1.510&4.607\\
    C I  &6713.586&-3.170&8.537\\
    Fe I &6713.745&-1.600&4.795\\
    \hline
    \end{tabular}
  \end{table}

To determine the lithium abundance $\log\epsilon$(Li), we calculated synthetic spectra in a region from 6703 to 6714~\AA\ and fitted the computed blend at 6707.8~\AA\ to the observed one. The list of spectral lines in the region is presented in Table~2; the line wavelengths, their oscillator strengths $\log gf$ and excitation potentials $E$ of the lower levels are shown (Yakovina et al. 2011). Note that the list is compiled on the basis of the VALD database (Kupka et al. 1999, Heiter et al. 2008; see also http://vald.astro.uu.se/$\sim$vald/php/vald.php); it is supplemented by ionic lines of rare earth elements from Shavrina et al. (2003) and from the DREAM database (Bi\'{e}mont et al. 1999). For three Fe~I lines (6703.567, 6705.101 and 6707.432) the $gf$-values are found by Yakovina et al.(2011) from comparison of their computed profiles with observed spectra of the Sun, Arcturus and some carbon stars. Data for molecular CN lines are taken from the Jor.-cor. list of Yakovina et al. (2011).

      It is important to note that we present in Table~2 only those lines which are dtectable in the synthetic spectra, especially in the immediate vicinity of the Li I 6708 line. However, the actual number of lines used in the computations is much larger. For instance, our full list contains about 150~CN lines within $\pm$1 \AA\ around the Li I 6708 line, but most of them are extremely weak and, therefore, are not shown in Table~2 (very weak C$_{2}$ lines from the list are not shown either). Moreover, Yakovina et al. (2011) presented in their Table~4 only a limited part of the Jor.-cor. list, which include the lines of special interest for their study of Li abundances in carbon stars. However, the full Jor.-cor. list contains many lines that are not presented there, but included in our computations (e.g., the CN 6708.972 \AA\ line and many others).

      Atomic data for the components of the $^7$Li line are taken from Lambert \& Sawyer (1984); note that the original source of the $gf$-values there was Gaupp et al.'s (1982) work. Later Smith et al. (1998) published the slightly corrected data for this line, which provide slight changes in the wavelengths and negligible changes in the $gf$-values; this update does not affect the derived Li abundances. It should be noted that we ignored the $^6$Li lines, because they are weak in computed spectra of programme stars.

      Detailed analysis of the Li I 6708 line demands careful consideration of blending lines. In particular, a few CN lines (see Table~2) can play a marked role for relatively cool G-type stars. As mentioned above, we used data for CN lines from the Jor.-cor. list of Yakovina et al. (2011). We used as well for comparison data for CN lines from the LJQ list of Yakovina et al. (2011); no marked changes in the derived Li abundances were found. The C and N abundances adopted by us in the CN line computations were generally equal to the solar ones (Asplund et. al. 2009). In a few cases, the C abundance was decreased and the N abundance was increased 
to better fit the CN lines in the observed spectrum.

      A Nd II line at 6708.03 \AA\ can contribute to the 6708 \AA\ blend. We used in computations of this line both the solar Nd abundance and the Nd abundance calibrated according to the individual Fe abundance for each star (Table~1). In both cases the computed Nd II line tends to be somewhat stronger than the observed one. Shavrina et al. (2003) supposed that the oscillator strength $\log gf$ adopted for this line is likely to be overestimated  by up to 0.3 dex. We found that the uncertainty in $\log gf$ of the Nd II line can lead to error of about 0.1 dex in the $\log \epsilon$(Li) upper limits, whereas for stars with the detectable Li abundances the error is much less.

% #3.2
\subsection{Computations of  lines and synthetic spectra}
      At the outset, LTE was assumed for all lines when fitting the synthetic spectra to the observed one; the lithium LTE abundance in a star is obtained as a result. Next, we determined the non-LTE correction for the Li~I~6708 line. For a set of abundances $\log\epsilon$(Li) and the star's model atmosphere, we calculate both LTE and non-LTE equivalent widths $W$; relations between $W$ and $\log\epsilon$(Li) (i.e. curves of growth) are constructed. Comparing the two relations, we obtain the non-LTE correction that corresponds to the derived LTE abundance. 
      (The non-LTE computations were implemented as well for the subordinate Li~I line at 6103.6~\AA; however, this line turned out to be too weak for detection in our stars.) 

      In our non-LTE analysis of the Li~I lines we utilized a 20-level Li model atom and followed the procedure described by Pavlenko \& Magazzu (1996) and Pavlenko et al. (1999). The adopted Li model atom allowed us to consider the interlocking of the Li~I lines (transitions) in an appropriate way. The Li ionization equilibrium is maintained by the whole system of the bound-free transitions. However, as noted by Pavlenko (1991), the transitions from/to the second level play the main role.  Overlapping absorption by atomic and molecular lines was considered.

      The radiation field in bound-free transitions is very important in non-LTE computations of the Li~I lines. In fact, the effectiveness of the Li over-ionization depends directly on the mean intensities of radiation field. When computing fluxes in the frequencies of bound-free transitions, we take into account absorption by atomic lines; moreover, we include the molecular line absorption in the background opacity. In our computations, we account for the multiplet structure of the lithium lines. 

      When computing synthetic spectra, we included the effect of instrumental broadening; following Tull et al. (1995), we adopted the value FWHM = 0.057 \AA\ for instrumental profile of the 2.7 m telescope spectrometer employed in our observations at the McDonald observatory. A more significant broadening is associated with the rotational and macroturbulent velocities; convolution with both these velocities was used in our computations of synthetic spectra.

% #3.3
\subsection{Computations of model atmospheres}

      Model atmospheres for programme stars are based on their parameters $T_{\rm eff}$, $\log g$, $V_t$ and $\log\epsilon$(Fe) presented in Table~1. In the model's computations we used the code SAM12 (Pavlenko, 2003), which is a modification of the code ATLAS12 of Kurucz (1999). It should be noted that Kurucz's (1993) code ATLAS9 was used for comparison for some stars. We found that the ATLAS9 models provide the same abundances $\log\epsilon$(Li) as the SAM12 models.  

      In the calculations of SAM12 model atmospheres, the abundances of 79 chemical elements (including hydrogen and helium) were specified as input parameters. So, we could calculate model atmospheres for stars with solar and non-solar chemical composition. The microturbulent parameter $V_t$ was assumed to be constant in a stellar atmosphere. Since the SAM12 code is constructed specially for studies of relatively cool stars, the presence of many molecules was taken into consideration. The system of equations for the molecular ionization-dissociation equilibrium was solved for a mixture of more than 100 constituents. The species and their ionization-dissociation equilibrium constants were chosen on the basis of our analysis of Tsuji's (1973) and Gurvitz et al. (1989) results. Convection was taken into account using a technique described in detail by Castelli et al. (1997). It should be noted that in atmospheres of cool supergiants, in contrast to cool dwarfs, convective transfer does not play a marked role. In particular, this effect seems to be unimportant for the analysis of all programme supergiants and giants including the coolest ones. 

      Some details concerning the opacity computations should be noted. In particular, we computed the cross-sections for bound-free absorption by C I, N I and O I atoms (Pavlenko \& Zhukovska 2003) using data from TOPBASE (Seaton et al. 1992). The opacity due to atomic and molecular absorption lines was included, using an opacity-sampling approach (Sneden et al. 1976). A list of atomic and molecular absorption lines in a region from 400 to 600,000~\AA\ was compiled from various sources, namely (i) atomic lines from VALD (Kupka et al. 1999, Heiter et al. 2008); (ii) relevant molecular lines from Kurucz's CDROM15 (Kurucz 1993); (iii) lines of H${_2}$O and OH  from Pavlenko (2003). The same opacity sources were adopted in computations of synthetic spectra with the code WITA6 (Pavlenko 1997). 

      We adopted  model atmospheres with a plane-parallel geometry. But for cool supergiants the effect of sphericity can play a marked role. This effect for stars with $T_{\rm eff}$ = 4000--6500 K has been investigated by Heiter \& Eriksson (2006); their general recommendation is that spherical models should be used for $\log g \leq$ 2 at $M$ = 1 M$_\odot$ and for $\log g \leq$ 1.5 at $M$ = 5 M$_\odot$. Most of our stars seem to be in a `safe range'. Ulrike Heiter (private communication) implemented test computations for the G-supergiant HR 8313 ($T_{\rm eff}$ = 4910 K; $\log g$ = 1.58), using the MARCS code (Gustafsson et al. 2008). She found that the maximum difference in the unconvolved flux in the considered spectral region between the plane-parallel and spherical models is less than 2 per cent. So, it is confirmed that the effect of sphericity can be ignored.

% #4
\section{DETERMINATION OF THE LITHIUM ABUNDANCES}

When spectra of cool supergiants and bright giants are studied, it is difficult to separate correctly the contributions of the projected rotational velocity $v \sin i$ and the macroturbulent velocity $V_{mac}$ to the line profiles. Therefore, we computed synthetic spectra in the 6708~\AA\ region separately for two limiting cases: (i) only with $v \sin i$; (ii) only with $V_{mac}$. We found that the derived Li abundances for these two cases are identical within 0.03 dex. (The distribution of macroturbulent velocities is assumed to be Gaussian.) 

      The rotational velocities $v \sin i$ for the programme stars were derived from profiles of rather strong Fe I lines in the vicinity of the Li I 6708 line.  The obtained $v \sin i$ values are presented in Table~1. For 40 of the 43 stars studied, the velocity $v \sin i$ is small, namely from 3.5 to 15 km~s$^{-1}$ with a mean value $v \sin i =$ 9.4$\pm$2.0 km~s$^{-1}$. For three stars we found higher velocities: $v \sin i =$ 25 km~s$^{-1}$ for HR~7542, 53 km~s$^{-1}$ for HD~17905 and 58 km~s$^{-1}$ for HR~7008 (the two latter stars are Li-rich giants, see below). It should be noted that all these $v \sin i$ evaluations are upper limits because we ignored $V_{mac}$ in the line profile fitting.  We have not attempted a simultaneous derivation of the rotational and macroturbulent velocities. 

      The lithium abundances $\log\epsilon$(Li), derived from the Li~I 6708 line both in LTE and non-LTE approach, are presented in Table~1 (two last columns). The first (upper) part of Table~1 contains the stars (11 in all) with an observable 6708 line and, therefore, with a detected Li abundance. Note that three objects there, namely HD~17905, HR~3102 and HR~7008, are Li-rich stars with $\log\epsilon$(Li)~=~3.2, 2.3 and 3.1, respectively. The second and lower part of Table~1 contains stars (32 in all), where the Li~I 6708 line is not seen, so only an upper limit for the Li abundance can be given. In order to evaluate an $\log\epsilon$(Li) upper limit, we varied $\log\epsilon$(Li) and searched for the value, which gives a Li~I 6708 line that is comparable with noise in the observed spectrum. Such a method agrees well with the supposition that the equivalent width $W$ of the minimum detectable Li~I 6708 line is 5 m\AA\ (see below). 

% Fig-1
\begin{figure}
\begin{center}
\includegraphics[width=83mm]{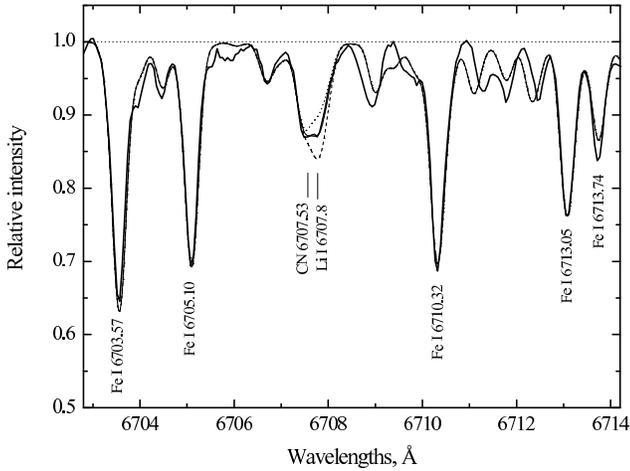}
\end{center}
\caption[]{Observed spectrum (thick solid curve) and synthetic ones in the Li I 6707.8 line region for the G5 supergiant HR 8313 (9~Peg). The Li~I 6707.8 line is computed for three Li abundances: the basic abundance $\log\epsilon$(Li) = 1.53 (thin solid curve) and two others corresponding to the uncertainty $\pm$0.14 dex (dashed and dotted curves). Apart from the Li~I line, positions of some other lines are shown.}
\end{figure}

We show in Fig.1 a fit of a synthetic spectrum to the observed one (thick solid curve) in a region around the Li~I 6707.8 line; the earlier-mentioned G5 Ib star HR~8313 (9~Peg) is chosen as an example. Computations of the 6707.8~\AA\ blend are performed for three Li abundances; namely, the abundance from Table~1  $\log\epsilon$(Li) = 1.53 (thin solid curve) and  for comparison computations corresponding to the uncertainty $\pm$0.14 dex (dashed and dotted curves). One sees that the best fit is really achieved at $\log\epsilon$(Li) = 1.53.

       For this relatively cool star, apart from the Li~I line, a CN line at 6707.53~\AA\ contributes to the 6707.8~\AA\ blend. To obtain the best fit, we adopted in Fig.1 the nitrogen abundance $\log\epsilon$(N) = 8.15 that is greater by 0.32 dex than the solar abundance $\log\epsilon_\odot$(N) =7.83 (Asplund et al. 2009). (Note that a similar or greater N excess is typical for other cool supergiants; see, e.g., Lyubimkov et al. 2011). The carbon abundance is adopted there to be solar. We varied the N and C abundances and found that an accurate fit to the 6707.8~\AA\ blend can be obtained for a greater N excess accompanied by a C deficiency, but this does not lead to a Li abundance change. The same N and C abundances describe well other CN lines in the Li~I 6707.8 line region. It should be noted that an accurate determination of the N and C abundances for programme stars is outside the present work.  

      Other lines are seen in Fig.1 (see Table~2 for details). The LTE assumption was used in the computation of these lines, of which the five strongest are Fe~I lines (marked in Fig.1). We tried to fit the Fe~I lines with a single Fe abundance; lines of other chemical elements were calculated with the solar abundances (Asplund et al. 2009). It is known that Fe~I lines in spectra of cool supergiants can be sensitive to departures from LTE and the Fe abundance may be underestimated using LTE analysis (see, e.g., Boyarchuk, Lyubimkov \& Sakhibullin, 1985). Our LTE fitting to the Fe~I profiles showed that the derived Fe abundances tend to be lower than the abundances found in Paper I from Fe~II lines, which are insensitive to the non-LTE effects. In particular, the coolest programme stars with temperatures $T_{\rm eff}$ from 4430 to 5000~K showed $\log\epsilon$(Fe) from Fe~I lines to be underestimated by 0.1-0.2 dex. In our computations of synthetic spectra we used the mean Fe abundance derived from Fe~I lines around the Li~I 6707.8 line. 

      One may see from Fig.1 that there are three weak lines in the observed spectrum to the right of the 6707.8 blend, which are poorly fit by the synthetic spectrum. In other words, there is a shift between their wavelengths in the observed and calculated spectra. We found that these lines belong to the molecule CN. We inspected a few  lists of the CN lines and noted that there are similar random shifts in the wavelengths of some weak CN lines in all the lists (they are different in different lists). However, positions of the CN lines in the immediate vicinity of the Li~I 6707.8 line are known with confidence, so they cannot contribute any uncertainty to the Li abundance derivation.  

% Fig-2
\begin{figure}
\begin{center}
\includegraphics[width=83mm]{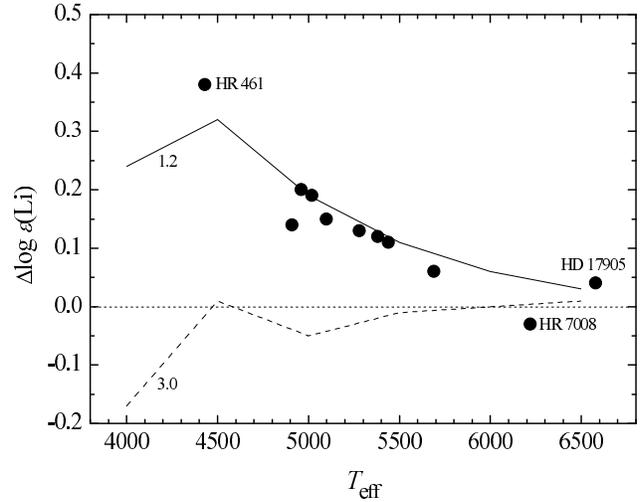}
\end{center}
\caption[]{Difference between the non-LTE and LTE lithium abundances for 11 stars with detectable $\log\epsilon$(Li) values from Table~1. For comparison we show the non-LTE corrections of Lind et al. (2009) computed for models with the LTE abundances $\log\epsilon$(Li) = 1.2 (solid line) and $\log\epsilon$(Li) = 3.0 (dashed line). See text for details.}
\end{figure}

      It is interesting to analyse the difference $\Delta\log\epsilon$(Li) between the non-LTE and LTE abundances $\log\epsilon$(Li) (see two last columns in Table~1). We show in Fig.2 this quantity as a function of the effective temperature $T_{\rm eff}$ for 11 stars with detected Li abundances. One sees that the difference is mostly positive. Moreover, there is an evident anti-correlation between $\Delta\log\epsilon$(Li) and $T_{\rm eff}$; so, on average, the cooler a star the greater $\Delta\log\epsilon$(Li). The smallest $\Delta\log\epsilon$(Li) values are found for the relatively hot F stars HR~7008 and HD~17905, both Li-rich (marked in Fig.2). These two $\Delta\log\epsilon$(Li) values are close to zero or even negative (HR~7008). The greatest non-LTE correction of 0.38 dex is found for the coolest supergiant of our sample, namely HR~461 (K0~Ia). 

      For comparison, we show in Fig.2 results of Lind et al.'s (2009) non-LTE computations for models with $\log g$ = 2.0, $[Fe/H]$ = 0.0, $V_t$ = 2.0 km~s$^{-1}$ and the LTE abundance $\log\epsilon$(Li) = 1.2 (solid line). Note that $\log\epsilon$(Li) = 1.2 corresponds to the mean LTE value for 8 stars with the moderate Li abundances from the upper part of Table~1 (excluding the three Li-rich stars). There is a good agreement between our non-LTE corrections and Lind et al.'s predictions. In particular, the correction 0.38 dex found for the coolest star HR~461 is virtually coincident with the value 0.41 dex presented by Lind et al. for a model with the very similar parameters $T_{\rm eff}$ = 4500~K, $\log g$ = 1.0, $V_t$ = 2.0 km~s$^{-1}$ and $\log\epsilon$(Li) = 0.9 (LTE). 

      We also show in Fig.2 Lind et al.'s computations for the higher LTE abundance $\log\epsilon$(Li) = 3.0 (dashed broken line). One sees that at $T_{\rm eff}$ = 6200-6600~K, where the Li-rich stars HR~7008 and HD~17905 are located, the difference between the cases $\log\epsilon$(Li) = 1.2 and $\log\epsilon$(Li) = 3.0 is small, but it increases significantly as $T_{\rm eff}$ decreases. For stars with $T_{\rm eff}$ = 4000-5000~K, the non-LTE correction depends markedly on the Li abundance. 

      It is necessary to discuss briefly the accuracy of the non-LTE lithium abundances presented in Table~1. For 11 stars with the definitely determined $\log\epsilon$(Li) values, specifically from 1.16 to 3.18 (upper part of Table~1), we show the errors of the $\log\epsilon$(Li) determination. One sees that they vary from $\pm$0.10 to $\pm$0.21dex with a mean of $\pm$0.15 dex. The tabulated errors incorporate four contributors, namely the uncertainty in the observed 6708 line profile and the uncertainties in the parameters $T_{\rm eff}$ , $\log g$ and $V_t$. The uncertainties in $T_{\rm eff}$ give the main contribution to the errors in $\log\epsilon$(Li). For stars with a $\log\epsilon$(Li) upper limits (lower part of Table~1), the typical errors in the limits are estimated to be 0.2-0.3 dex. 

     It is known that in spectra of some Li-rich stars the weaker subordinate Li~I line at 6103.6~\AA\ is observable. We tried to find this line in spectra of our three Li-rich stars, namely HD~17905, HR~3102 and HR~7008. Unfortunately, the line is not discernible there. Computations of synthetic spectra in the 6103.6~\AA\ region with the corresponding $\log\epsilon$(Li) values confirmed that the line is really too weak (note as well that the line is located in the wing of a rather strong blend). Moreover, we tried to find the Li~I 6103.6 line in a spectra of the coolest star in our list, namely HR~461 ($T_{\rm eff}$ = 4430~K; $\log\epsilon$(Li) = 1.27). We obtained the same result, i.e., the line is not seen and computations confirm its weakness. An inspection of spectra of all other stars showed that the Li~I 6103.6 line is not detectable there, too, because of the rather low Li abundances.

% #5
\section{COMPARISON OF LITHIUM ABUNDANCES WITH PUBLISHED DATA}

When inspecting the derived Li abundances in Table~1, we may divide the programme stars into three groups, namely:

   (i) Three Li-rich F-type stars, i.e. HR~3102, HR~7008 and HD~17905 with $\log\epsilon$(Li)~=~2.3, 3.1 and 3.2, respectively. 

   (ii) Eight more stars (all the G-type ones) with detectable Li abundances. It is interesting that these abundances occupy a rather narrow range $\log\epsilon$(Li) from 1.2 to 1.5 with the mean 1.38$\pm$0.15 (the mean LTE abundance is 1.20). We shall show below that there is a good explanation in the theory for most of the group. 

   (iii) The remaining 32 stars with  $\log\epsilon$(Li) upper limits (lower part of Table~1).
     
      One sees that only for 25 per cent of programme stars are measurable $\log\epsilon$(Li) values found. Therefore, most F- and G-type supergiants and giants do not show a detectable Li abundance. This conclusion agrees well with data of other authors (see, e.g., Gonzalez et al. 2009, where only for 13 of 417 G and K giants and supergiants in the Galactic bulge, i.e., 3 per cent, was a detectable Li~I 6708 line found). 

      A comparison with published Li abundances accomplishes two purposes. First, it provides a check on our and published abundances. We believe that in the present era of digital spectra and abundance analyses using model atmospheres and synthetic spectra, such an abundance comparison is effectively
 a comparison of the adopted effective temperatures $T_{\rm eff}$ (if the same approach, LTE or non-LTE, is applied). In fact, due to the strong sensitivity of the Li~I~6708 line to $T_{\rm eff}$, the difference in $T_{\rm eff}$ is really the  dominant source of differences in $\log\epsilon$(Li). Other possible sources, e.g. differences in methods of computations of model atmospheres and synthetic spectra, are of minor importance. (As a rule, when the temperatures $T_{\rm eff}$ for a star in two compared works are close, the Li abundances are close, too).  Second, a critical analysis of previous results provides an opportunity to enlarge the sample of bright giants and supergiants with a known Li abundance by adding of published (and adjusted) results to our data. 

% Fig-3
\begin{figure}
\begin{center}
\includegraphics[width=83mm]{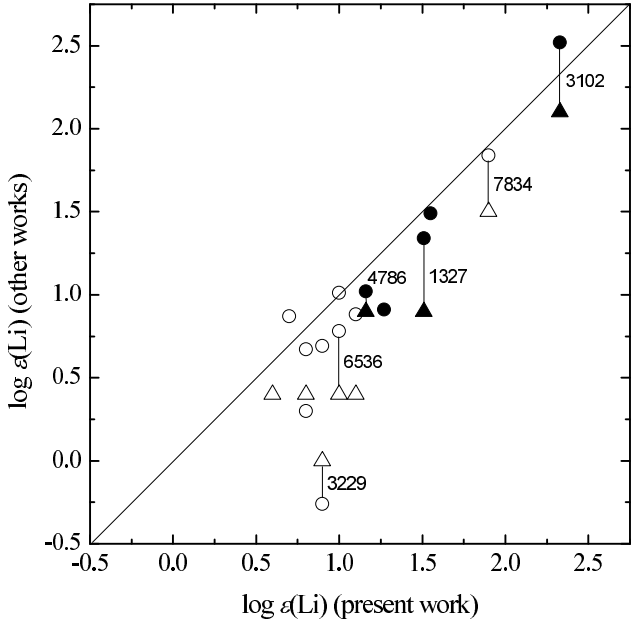}
\end{center}
\caption[]{Comparison of our non-LTE $\log\epsilon$(Li) values with data from other works. Circles correspond to common stars from 'Luck's group', triangles correspond to common stars from 'L\`{e}bre's group' (see text for details). The stars with the detectable Li abundances are displayed by filled symbols, and the stars with the Li upper limits are shown by open symbols. HR numbers of some stars are marked.}
\end{figure}

      We compare our results for $\log\epsilon$(Li) with two groups of published data for local supergiants and giants, namely (i) the non-LTE lithium abundances of Luck (1977) for G and K supergiants and the LTE abundances of Luck \& Wepfer (1995) for F and G bright giants; Luck \& Heiter's (2007) LTE abundance for the giant HR~1327 is used, too; (ii) the LTE abundances of L\`{e}bre et al. (2006) for F, G and K bright giants. The comparison is shown in Fig.3. Published results for less luminous or less massive stars are not considered here; see, for example, Wallerstein et al.'s (1994) analysis of F2-G5 giants (luminosity class III) in the Hertzsprung gap. 

      Common stars from Luck's group are shown by circles and from L\`{e}bre's group by triangles. Stars with detectable Li abundances (filled symbols) and the stars with the Li upper limits (open symbols) are presented. This figure shows that our non-LTE values $\log\epsilon$(Li) are in good agreement with the $\log\epsilon$(Li) values (one half is LTE, another half is non-LTE) for common stars of Luck's group. On the other hand, the LTE data for common stars of L\`{e}bre's group tend to show lower Li abundances as compared with our non-LTE results. We suppose that this discrepancy is mostly explained by the neglect of departures from LTE by L\`{e}bre et al. 

      The bright giant HR~3229 (G5 II) is an evident 'outlier' in Fig.3: both Luck \& Wepfer (1995) and L\`{e}bre et al. (2006) give a much lower Li upper limit for this star than our evaluation. A likely explanation may be connected with differences in the adopted effective temperatures $T_{\rm eff}$. The temperatures $T_{\rm eff}$ = 4725 and 4700~K adopted in the published papers are markedly cooler than our value $T_{\rm eff}$ = 5130~K. Such an underestimation of $T_{\rm eff}$ can lead to a 0.4 dex decrease of the Li upper limit. As far as the $T_{\rm eff}$ determination for HR~3229 is concerned, one may mention Giridhar et al.'s (1997) value $T_{\rm eff}$ = 5000~K that is in good agreement with our estimation. It should be noted as well that our value $T_{\rm eff}$ = 5130~K agrees very well with the mean effective temperature $T_{\rm eff}$ = 5120~K for G5 II stars, which follows from our $T_{\rm eff}$ scale (see Table~4 in Paper I). 

      Now we consider in more detail results for the three Li-rich stars. In particular, two interesting objects, namely HD~17905 and HR~7008, were not included in Paper I. The Li-rich supergiant HD~17905 was discovered by Kovtyukh et al. (2005). They found the following parameters for this star: $T_{\rm eff}$ = 6407~K, $\log g$ = 2.3 and $\log\epsilon$(Li) = 3.15. We redetermined the $T_{\rm eff}$ and $\log g$ values using our method, which is based on the $\log g$ derivation from stellar parallaxes (Lyubimkov et al. 2009, 2010). The parallax of the star is $\pi$ = 6.36$\pm$0.47 mas (van Leeuwen 2007); it provides the star's distance $d$ = 157$\pm$12 pc. As mentioned above, the parallax is a good indicator of $\log g$. The [$c_1$] index in the uvby photometric system was used by us as a $T_{\rm eff}$ indicator; [$c_1$] = 0.673 for HD~17905 (Hauck \& Mermilliod 1998). The interstellar extinction $A_{\rm v}$ is also needed for the $\log g$ determination; we found $A_{\rm v}$ = 0.17. Our final parameters for HD~17905 are the following: $T_{\rm eff}$ = 6580$\pm$100~K, $\log g$ = 3.26$\pm$0.05 and $\log\epsilon$(Li) = 3.18; the projected rotational velocity is $v \sin i$ = 53 km~s$^{-1}$. Note that our $\log g$ value is significantly greater than Kovtyukh et al.'s value $\log g$ = 2.3 and corresponds not to a supergiant, but to a giant (luminosity class III). However, since the Li abundance is weakly dependent on $\log g$, our and Kovtyukh et al.'s abundances are very close.

      The Li-rich giant HR~7008 (HD~172365) was also studied by Kovtyukh et al. (2005) who determined the following parameters: $T_{\rm eff}$ = 6030~K, $\log g$ = 2.5 and $\log\epsilon$(Li) = 3.12. The parallax of the star is $\pi$ = 2.92$\pm$0.54 mas (van Leeuwen 2007); so its distance is $d$ = 342$\pm$12 pc. The index [$c_1$] = 0.622 follows from Hauck \& Mermilliod's (1998) catalogue. Using our evaluation of the interstellar extinction ($A_{\rm v}$ = 0.51) we obtained that $T_{\rm eff}$ = 6220$\pm$100~K, $\log g$ = 2.53$\pm$0.11 and $\log\epsilon$(Li) = 3.07; the projected rotational velocity is found to be $v \sin i$ = 58 km~s$^{-1}$. The star's parameters $T_{\rm eff}$ and $\log g$ determined in the two works are close, so the derived Li abundances are very close, too. 

% Fig-4
\begin{figure}
\begin{center}
\includegraphics[width=83mm]{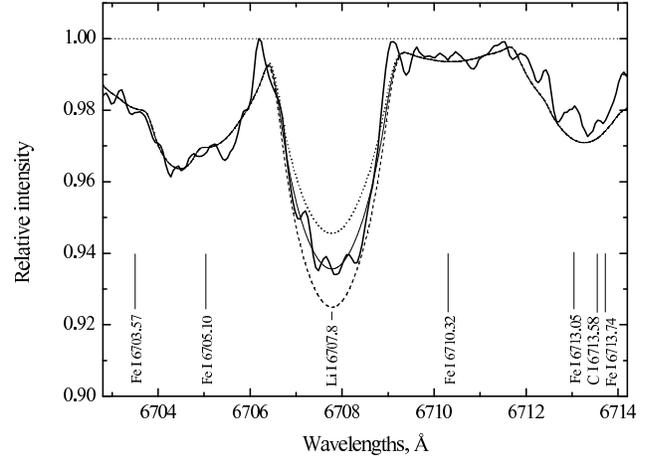}
\end{center}
\caption[]{Observed spectrum (thick solid curve) and synthetic ones in the Li~I 6707.8 line region for the Li-rich F-type giant HR~7008 with the relatively high rotational velocity $v \sin i$ = 58 km~s$^{-1}$. The Li~I 6707.8 line is computed for three Li abundances: the basic abundance $\log\epsilon$(Li) = 3.07 (thin solid curve) and two others corresponding to the uncertainty $\pm$0.12 dex (dashed and dotted curves).}
\end{figure}

It should be noted that in the determination of the interstellar extinction $A_{V}$ for these two new objects (HD~17905 and HR 7008), and  all stars in Paper I, we applied the method used in our study of a large sample of B-type stars (Lyubimkov et al. 2002). Using the $uvby$ photometric data and Jordi et al.'s (1997) code, we found the colour excess $E(b-y)$ and then $A_{V}$ through relation $A_{V} = 4.3 E(b-y)$. We concluded in Paper I that `uncertainties in $A_{V}$ affect slightly the surface gravities $\log g$ derived from parallaxes'.

The giant HR~7008 seems to have the greatest rotational velocity in our sample. Fig.4 compares the synthetic and the observed spectra of this star in the region of the Li I 6707.8 line. Computations of the 6707.8 \AA\ blend are performed for three Li abundances; namely, the value from Table~1 ($\log\epsilon$(Li) = 3.07 (thin solid curve)), and values corresponding to the uncertainty $\pm$0.12 dex (dashed and dotted curves).

Because of the rapid rotation, wide blends are seen there instead of sharp lines (compare with Fig.1 for the star HR~8313 with $v \sin i$ = 10 km~s$^{-1}$). Two blends in Fig.4, left and right of the lithium line, are formed mainly by Fe I lines. The left blend contains the Fe I lines at 6703.57 and 6705.10~\AA; the right one includes the weaker Fe I lines at 6713.05 and 6713.74~\AA. Moreover, the latter blend contains the C~I line at 6713.58~\AA, which gives a marked contribution for stars with $T_{\rm eff} >$~6000~K. We decreased the carbon abundance by 0.6 dex to obtain a better fit to the observed blend. As known, a carbon deficiency is observed in many F and G giants and supergiants. Note that the C reduction by 0.6 dex does not affect the derived high Li abundance. 

      The third Li-rich giant in Table~1 is HR~3102. Luck \& Wepfer (1995) obtained for this star $\log\epsilon$(Li) = 2.52, whereas L\`{e}bre et al. (2006) provided $\log\epsilon$(Li) = 2.1. Our value $\log\epsilon$(Li) = 2.33 is approximately the mean of these preceding values. Although the two earlier $\log\epsilon$(Li) evaluations for HR~3102 are based on a LTE approach, the non-LTE correction for this F7-type star is small (0.06 dex, see Table~1). Even smaller non-LTE corrections are needed for the  LTE abundances of Kovtyukh et al. (2005) for the Li-rich giants HD~17905 and HR~7008. 

      Summarizing, we may say that our Li abundances for common stars and, in particular, for the three Li-rich stars are in good agreement with published results.  This demonstration encourages us to add some of the published results to ours to increase the sample size. When adding data of other authors, it is necessary to redetermine their parameters $T_{\rm eff}$ and $\log g$ with our method and then to correct the Li abundance. We found that Luck \& Wepfer's (1995) data are quite suitable for a significant enlargement of the group of stars with detected Li abundances.

% #6
\section{ADDITIONAL STARS FROM LUCK \& WEPFER (1995) LIST}

Luck \& Wepfer (1995, hereinafter LW'95) studied 38 F- and G-type stars, which have been classified as bright giants (luminosity class II or II-III) in the Catalogue of Bright Stars (Hoffleit \& Jaschek 1982). Note that this old classification seems to be erroneous for some of the stars (see below). In LW'95, the resonance Li~I 6708 line, as well as the spectrum synthesis was used for the LTE Li abundance analysis. One may see from LW'95 that the detectable Li abundances were derived for the majority of the stars. However, they presented for some stars  very low abundances, $\log\epsilon$(Li) between -0.3 and 0.3, with correspondingly  very small measured equivalent widths, $W$(6708) = 1-5 m\AA\ (see Table~10 in LW'95). Note that our LTE upper limits on the Li abundance are not lower than 0.5 (Table~1); they correspond on average to a minimum detectable equivalent width $W$(6708) = 5 m\AA\ (see below). We believe that such low Li abundances in LW'95 are questionable. Therefore, a careful selection of LW'95 stars is needed before including them in further analysis.

     In order to elevate a reliability of our abundance determinations for LW'95 stars, we implemented a severe selection of objects from LW'95's list. Stars with very small equivalent widths $W$(6708) and/or very low Li abundances were excluded. Only stars with reliably detected Li abundances were included. Specifically, two requirements were imposed: i) $W$(6708) $>$ 6 m\AA; ii) $\log\epsilon$(Li) $>$ 0.7. Next, we selected those stars with accurate values of the parallax $\pi$ and photometric index [$c_1$] in order to apply with confidence our method of the $T_{\rm eff}$ and $\log g$ determination.

      When parameters $T_{\rm eff}$ and $\log g$ are redetermined, spectral subtypes and luminosity classes can be updated. We found that the old classification used in LW'95 is frequently erroneous;  some stars seem not to be bright giants. For instance, for the star HR~8718 classified as F5 II we derived with very high accuracy $T_{\rm eff}$ = 6570 and $\log g$ = 4.01 (from $\pi$, [$c_1$] and $\beta$-index). According to the derived $\log g$ value, this nearby star ($\pi$ = 25.66 mas, so $d$ = 39 pc) is a dwarf (luminosity class V), not a bright giant. So, we excluded HR~8718 from further analysis. Moreover, four stars selected have $\log g$ = 3.4-3.6 that are typical for the luminosity class III. These four giants are interesting, because (i) they all are Li-rich stars and (ii) they all have virtually the same mass $M \approx$ 2 M$_\odot$, so they supplement markedly the low-mass end of our sample. Therefore, we retained these four stars in the list. 

% Table-3
\begin{table*}
 \centering
 \begin{minipage}{150mm}
 \caption{Additional 12 stars from Luck \& Wepfer's (1995) list. Upper part of the table - bright giants 
(luminosity class II or II-III), lower part - giants (luminosity class III). In two last columns the lithium 
abundances are presented, namely LW'95 original LTE values and our corrected non-LTE values.}
    \begin{tabular}{ccccccccccc}
    \hline
 HR & HD & Sp$^\ast)$ & $v \sin i$, & $\pi$, mas & \textit{d}, pc & $T_{\rm eff}$ & $\log g$ & $M/M_{\odot}$ & $\log \epsilon$(Li) & $\log\epsilon$(Li)\\
    &    &            &km~s$^{-1}$  &            &                &               &          &               & (LW'95)             &(corrected)\\
    \hline
      59&1227  &G1 II-III&$<$ 1  & 7.99$\pm$0.60 &125&5330&2.95&2.74&0.76&1.12\\
    1135&23230 &F3 II    &48.9   & 5.87$\pm$0.22 &170&6570&2.39&5.01&2.29&2.21\\
    1644&32655 &A9 II-III&-      & 4.23$\pm$0.76 &236&7150&2.98&3.30&2.21&2.36\\
    3182&67447 &G5 II    &4.4    & 4.35$\pm$0.26 &230&5110&2.23&4.54&1.12&1.64\\
    3306&71115 &G1 II-III&$<$ 1.2& 8.26$\pm$0.32 &121&5360&2.67&3.34&0.80&1.23\\
    3557&76494 &G8 II    &4.6    & 2.30$\pm$0.58 &435&5040&2.00&5.30&1.00&1.36\\
    6707&164136&F5 II    &-      & 3.79$\pm$0.39 &264&6410&2.29&5.31&2.05&1.95\\
    7921&197177&G0 II    &1.7    & 4.06$\pm$0.41 &246&5500&2.37&4.23&1.37&1.83\\
    \hline
    1298&26574 &F0 III   &-      &26.80$\pm$0.32 &37 &7040&3.62&2.12&2.81&2.76\\
    2936&61295 &F1 III   &34.9   & 9.84$\pm$0.47 &102&6880&3.51&2.10&3.04&2.88\\
    5913&142357&F4 III   &27.4   &10.55$\pm$0.50 &95 &6530&3.48&2.00&2.21&2.27\\
    6604&161149&F1 III   &65.8   & 8.33$\pm$0.40 &120&6910&3.42&2.24&2.26&2.42\\
    \hline
    \end{tabular}
$^\ast$) Spectral classification is corrected in accordance with the derived parameters $T_{\rm eff}$ and $\log g$ (see text for details). 
  \end{minipage}
\end{table*}

      The 12 selected LW'95 stars are presented in Table~3. The four above-mentioned giants are shown separately in the lower part of the table. Apart from HR and HD numbers and the corrected spectral classification, we give there the projected rotational velocity $v \sin i$, where it is known (de Medeiros \& Mayor 1999). Stellar parallaxes $\pi$ are taken again from van Leeuwen's (2007); the corresponding distances $d$ are found to be less than 270 pc excluding the star HR~3557 with $d$ = 435 pc. Parameters $T_{\rm eff}$ and $\log g$ are determined by our method (Paper I) using the parallax $\pi$ and the photometric index [$c_1$]; the latter value is found with Hauck \& Mermilliod's (1998) catalogue. (As noted in Paper I, the typical errors in our parameters $T_{\rm eff}$ and $\log g$ for such nearby stars are $\pm$120~K and $\pm$0.06~dex, respectively). Moreover, the interstellar extinction $A_{\rm v}$ is needed; it is determined from the relation $A_{\rm v}$ = 3.12 $E(B-V)$, where $E(B-V)$ is given in LW'95. Given the derived parameters $T_{\rm eff}$ and $\log g$, we find the masses $M$ of the stars from Claret's (2004) evolutionary tracks. One may see from Table~3 that all the stars have the masses $M <$~6~M$_\odot$. Note as well that the former classification of the stars used by LW'95 is corrected in accordance with the new values $T_{\rm eff}$ and $\log g$. The updated spectral subtypes and luminosity classes presented in Table~3 are obtained with the help of Paper I for bright giants and Alonso et al.'s (1999) $T_{\rm eff}$ scale for giants.  

      For the 12 selected stars, we corrected LTE lithium abundances obtained in LW'95. The first correction accounts for  differences the parameters $T_{\rm eff}$ and $\log g$. The measured equivalent widths $W$(6708) of the Li~I line presented in LW'95 are used in order to obtain $\log\epsilon$(Li) for LW'95 parameters and then for our  values. Difference between our and LW'95 values of $T_{\rm eff}$ for the 12 stars varies from -170 to 350~K that leads to the corrections in the Li abundances from -0.16 to 0.37 dex. Since the equivalent widths $W$(6708) for the stars are rather small and, therefore, the observed line Li~I 6708 is insensitive to the microturbulent parameter $V_t$, we adopted in our computations the $V_t$ values from LW'95. 

      The second correction concerns the transition from LTE to non-LTE Li abundances. We applied Lind et al.'s (2009) non-LTE corrections tabulated in dependence on $T_{\rm eff}$, $\log g$ and lithium LTE abundances. Using the abundances $\log\epsilon$(Li) obtained after the first correction, we interpolated  in Lind et al.'s data on $T_{\rm eff}$, $\log g$ and $\log\epsilon$(Li). In the two last columns of Table~3, we present both the original LW'95 lithium LTE abundances and our corrected non-LTE values.  

      Thus, critical analysis of LW'95 data allows us to increase a total number of stars in our sample from 43 to 55 and, in particular, to enlarge significantly our group of the stars with the detected Li abundances. Altogether we have now 23 supergiants and giants with definite $\log\epsilon$(Li) values. One sees from Table~3 that 7 of 12 additional stars are Li-rich (if the star HR~6707 with $\log\epsilon$(Li) = 1.95 is included), so the total number of Li-rich stars is 10 now. One will see below that the Li abundances found for the stars of additional sample (Table~3) are in good agreement with conclusions obtained for the stars of main sample (Table~1).

% #7
\section{LITHIUM ABUNDANCE AS A FUNCTION OF EFFECTIVE TEMPERATURE $T\lowercase{_{\rm eff}}$}

Relations between the lithium abundance $\log\epsilon$(Li) and two fundamental parameters of the stars, namely their effective temperature $T_{\rm eff}$ and mass $M$, are of special interest, especially as regards a star's location in the H-R diagram. Those supergiants and giants, which have crossed the Hertzsprung gap and experienced the first dredge-up phase, will have their surface Li abundance lowered by a factor of about 50 (see, e.g., Iben 1967). In the event that Li was not destroyed or depleted  prior to the onset of the first dredge-up (i.e., during the MS phase), the maximum Li abundance in a post-first-dredge-up star will be $\log\epsilon$(Li) of about 1.5 dex. The second dredge-up at the termination of the red-blue loops will result in a further lowering of the Li abundance. And, of course, destruction of Li in the MS progenitor will lower the Li abundance also. In a simple picture, these ideas suggest that a Li-rich star as defined by a Li abundance $\log\epsilon$(Li) $\ge$ 2.0 may be either a star that did not deplete Li as a MS star and has not yet to experience the first dredge-up, or if it experienced the first dredge-up must have synthesized Li. Location in the H-R diagram offers a clue as to the likely interpretation, but a decision must be consistent with other abundance data, particularly the C and N abundances. 

% Fig-5
\begin{figure}
\begin{center}
\includegraphics[width=83mm]{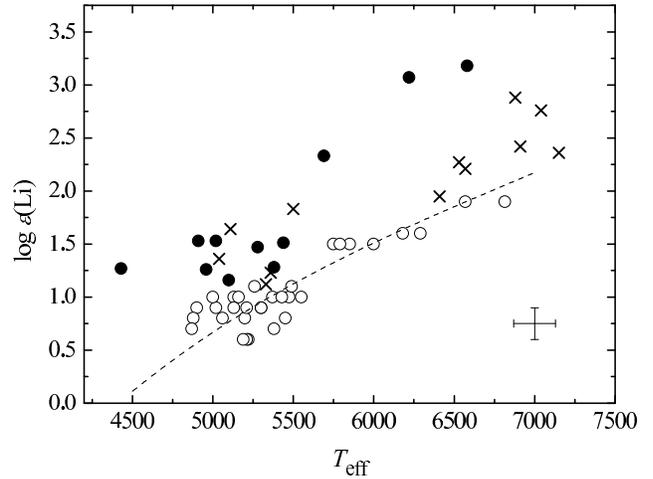}
\end{center}
\caption[]{Lithium abundance $\log\epsilon$(Li) as a function of the effective temperature $T_{\rm eff}$. Filled circles correspond to 11 stars with detected $\log\epsilon$(Li) from Table~1. Open circles show the stars with Li upper limits ibidem. Crosses present additional LW'95 stars from Table~3. Dashed line is computed on the supposition that the equivalent width $W$ of the minimum detectable Li~I 6708 line is 5 m\AA. Typical error bars are shown right below for $T_{\rm eff}$ and $\log\epsilon$(Li) ($\pm$130 K and $\pm$0.15 dex, respectively), that are mean values for the stars with detected lithium.}
\end{figure}

% Fig-6
\begin{figure}
\begin{center}
\includegraphics[width=83mm]{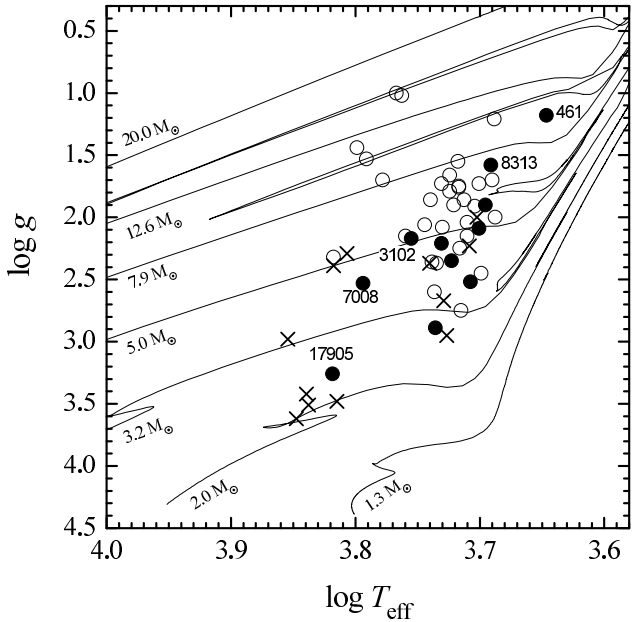}
\end{center}
\caption[]{Claret's (2004) evolutionary tracks and positions of programme stars on the $\log T_{\rm eff} - \log g$ plane. Symbols of the stars are the same as in Fig.5. Several stars of special interest are labeled by their HR numbers (see text).}
\end{figure}

      Before confronting observed Li abundances with theoretical predictions, it is helpful to make the obvious point that Li is observable only as a neutral atom, an alkali element with a low ionization potential and a low abundance. Thus, it is detectable spectroscopically only in cool stars and not detectable in the hotter B stars, i.e. the MS progenitors of F-G supergiants and bright giants. Therefore, processes affecting the surface abundance of Li in the B-type MS stars are hidden from spectroscopic scrutiny.

      We show in Fig.5 the lithium abundance $\log\epsilon$(Li) as a function of the effective temperature $T_{\rm eff}$ for all 55 stars studied. Filled circles correspond to the 11 stars with the detected Li abundances from Table~1; open circles represent the 32 stars with Li upper limits ibidem; crosses show the additional 12 stars from Table~3. It is interesting that the latter stars are distinctly divided into two groups, namely relatively cool G giants and hotter F ones. 

      One sees that there is a correlation for all the stars. Such a correlation for supergiants and bright giants is known from published work; see, e.g., Luck (1977), Luck \& Wepfer (1995), L\`{e}bre et al. (2006) and Gonzalez et al. (2009). The correlation is trivially explainable in the case of stars with Li upper limits (open circles in Fig.5); it reflects the
 strong dependence of the Li~I 6708 line on $T_{\rm eff}$. A given upper limit to the equivalent width of the 6708\AA\ line corresponds to a Li abundance limit that decreases with lower temperatures. The open circles in Fig.5 are well approximated by the dashed curve, which was computed on the supposition that the equivalent width $W$ of the minimum detectable Li~I 6708 line is 5 m\AA\ (we adopted $\log g$ = 2.0 for these non-LTE computations). 
Note that the detection limit $W$ = 5 m\AA\ agrees well on average with the minimum detectable values $W$ = 3 m\AA\ and $W$ = 7 m\AA\ adopted by Luck (1977) for cool supergiants with $T_{\rm eff} \ge$ 5000~K and $T_{\rm eff} \le$ 4600~K, respectively. Detection limits are necessarily greater for stars showing rapid rotation. 

      An apparent correlation between $\log\epsilon$(Li) and $T_{\rm eff}$ in the case of the stars with the detected $\log\epsilon$(Li) values is more difficult to explain. It is interesting that the relation between $\log\epsilon$(Li) and $T_{\rm eff}$ for them is not smooth. When considering the region $T_{\rm eff} <$ 5500~K, i.e., G-type supergiants and giants, one sees an absence of Li-rich stars. Moreover, eight G stars from Table~1 with $T_{\rm eff} <$ 5500~K (filled circles in Fig.5) show Li abundances within the narrow range $\log\epsilon$(Li)~= 1.2 to 1.6 (the mean is about 1.4), i.e., they show no correlation with $T_{\rm eff}$. The dashed curve in Fig.5 shows that abundances $\log\epsilon$(Li) = 1.2-1.5 are measureable for F stars but such Li abundances are too low to be detectable at $T_{\rm eff} \ge$ 6000~K. At $T_{\rm eff} \ge$ 6000~K, only Li-rich stars with $\log\epsilon$(Li) $\ge$ 2.0 are present (if stars with Li upper limits are not considered). Therefore, the $\log\epsilon$(Li) distribution with $T_{\rm eff}$ shows an apparent step at $T_{\rm eff}$ = 5500~K. It should be noted that a similar one-step distribution of detected Li abundances with $T_{\rm eff}$ has been obtained for F and G bright giants by Luck \& Wepfer (1995, see their Fig.21), as well as by L\`{e}bre et al. (2006, Fig.2). We shall show below that all Li-rich stars have rather low masses: $M <$~6~M$_\odot$; it is interesting that at the same time they occur only among relatively hot F-type stars.

% #8
\section{LITHIUM ABUNDANCE AS A FUNCTION OF MASS $M/M_\odot$}
% #8.1
\subsection{Observed relation between $\log\epsilon$(Li) and $M/M_\odot$}

 Stellar masses $M/M_\odot$ are derived by us from Claret's (2004) evolutionary tracks; in Fig.6 we placed the selected tracks in the log $T_{\rm eff}$-$\log g$ plane together with all 55 stars studied. As known, the existence of red-blue loops is a specific feature of the evolutionary diagram of F and G supergiants. One may see from Fig.6 that the loops are pronounced for stars with masses $M >$~6~M$_\odot$, but are undeveloped for stars with $M <$~6~M$_\odot$. In fact, Claret's (2004) computations show, on the one hand, that stars with $M$~=~3-5~M$_\odot$ after the beginning of the first dredge-up and subsequently during the whole red-blue loop phase are located in the narrow $T_{\rm eff}$ range between 4000 to 5000~K, i.e., they are K-type giants or supergiants. On the other hand, a star with $M$~=~6.3~M$_\odot$ during the loop phase reaches an effective temperature $T_{\rm eff}$ = 6600~K, i.e. such a K-type supergiant becomes a G- and F-type supergiant. Therefore, theory suggests there is a boundary at $M \approx$~6~M$_\odot$. We shall show below that a separation of the stars into two parts, namely $M <$~6~M$_\odot$ and $M >$~6~M$_\odot$, is of great significance from the viewpoint of both the derived Li abundances and their explanations. 

% Fig-7
\begin{figure}
\begin{center}
\includegraphics[width=83mm]{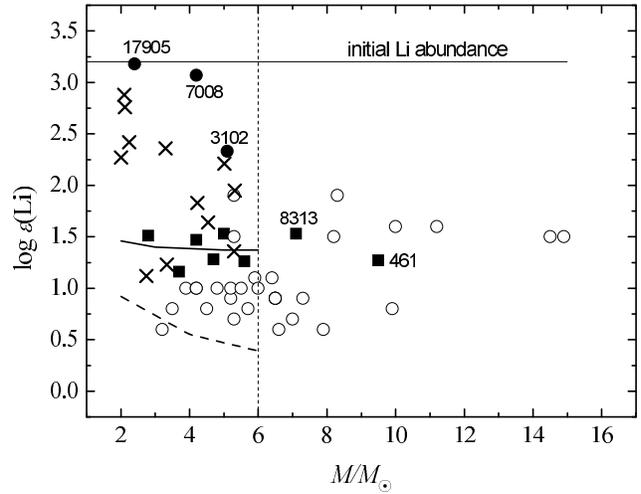}
\end{center}
\caption[]{The observed $\log\epsilon$(Li) $- M/M_\odot$ relation in comparison with theoretical predictions followed from model computations by Frischknecht (priv. comm.). Filled circles, filled squares and open circles correspond to three Li-rich giants, further eight stars with $\log\epsilon$(Li) = 1.2-1.6 and remaining stars with Li upper limits from Table~1, respectively. Crosses present additional LW'95 stars from Table~3. The initial Li abundance $\log\epsilon$(Li) = 3.2 is shown by thin straight line. Thick line in the $M$ = 2-6~M$_\odot$ range displays the predicted post-FD abundances for non-rotating models ($v_0$ = 0 km~s$^{-1}$); dashed line shows the predicted post-FD abundances for models with $v_0 \approx$ 50 km~s$^{-1}$.}
\end{figure}

      In Fig.7, relations between the Li abundances and the masses $M/M_\odot$ are shown: the thick solid line  and the dashed one correspond to theoretical predictions discussed below. One sees that for stars with $M\lesssim 6$ M$_\odot$ there is  a wide spread of $\log\epsilon$(Li) abundances. Just in this region are situated all 10 Li-rich stars with $\log\epsilon$(Li) from 2.0 to 3.2. Next, there are 11 more stars (6 from Table~1 - filled squares, and 5 from Table~3 -- crosses) with $\log\epsilon$(Li) values in the rather narrow range 1.1-1.8.  The remaining 13 stars (open circles) have Li upper limits down to 0.6 dex. 

      On the other hand, all stars with $M >$~6~M$_\odot$ are seriously depleted in Li. There are only two stars in this region, namely HR~8313 and HR~461 (marked in Fig.7), with detected $\log\epsilon$(Li). They are rather massive stars ($M$ = 7.1 and 9.5 M$_\odot$ respectively), so they may be located on the loops (see Fig.6). If they are evolving along the loops, they may have terminated the first dredge-up (FD) phase, i.e. they are post-FD objects. Note that the lifetime of massive F and G supergiants on the loops is much longer than the  lifetime as post-MS and pre-FD objects (see, e.g., Lyubimkov 1998, Table~1). One may see from Fig.7 that there are no Li-rich stars with $M >$~6~M$_\odot$.

% #8.2
\subsection{Masses of Li-rich stars}

The Li-rich stars with $\log\epsilon$(Li) $\ge$ 2.0 are of special interest. There are three such objects in Table~1, namely the F-type giants HD~17905, HR~7008 and HR~3102. For two of them, HD~17905 and HR~7008, we found $\log\epsilon$(Li) = 3.18$\pm$0.11 and 3.07$\pm$0.12, respectively; so, their Li abundances coincide with the above-mentioned initial abundance in young stars $\log\epsilon$(Li) = 3.2$\pm$0.1. The third Li-rich star, HR~3102, has $\log\epsilon$(Li)~=~2.3$\pm$0.2; therefore, its Li abundance was somewhat depleted during evolution. There are 7 more additional Li-rich stars with $\log\epsilon$(Li) = 2.0-2.9 in Table~3. Therefore, there are altogether 10 Li-rich stars in our sample. According to Tables~1 and 3, their masses $M$ vary from 2.0 to 5.3 M$_\odot$. It would be interesting to inspect for comparison the masses of Li-rich giants and supergiants in other studies. 

      Brown et al. (1989) evaluated the Li abundance for 644 G and K giants and supergiants; they found six Li-rich stars with $\log\epsilon$(Li) = 2.0-2.7. We derived the masses of these stars from Claret's (2004) evolutionary tracks and obtained a spread of $M$ from 1.7 to 3.4 M$_\odot$. When adding three more stars with $\log\epsilon$(Li) = 1.8-1.9 from their sample, we inferred for them $M$ = 2.5-5.2 M$_\odot$. Charbonnel \& Balachandran (2000) collected data on the Li abundances in 28 giants with $T_{\rm eff}$ = 3900-5500~K; 17 of them have the abundances $\log\epsilon$(Li) $\ge$ 2.0, i.e. they are Li-rich stars. Masses of these 17 stars are mostly known; they span the range from 1.1 to 4.7 M$_\odot$. Reddy \& Lambert (2005) and Kumar \& Reddy (2009) studied three super Li-rich bright K giants with $\log\epsilon$(Li) = 3.7-4.0. When deriving the masses of these stars from Claret's tracks, we obtained $M$ = 1.0-4.1 M$_\odot$. There are five Li-rich stars with $\log\epsilon$(Li) = 2.1-3.1 among 145 FGK bright giants analyzed by L\`{e}bre et al. (2006). Their masses determined by us from Claret's tracks are $M$ = 1.9-3.1 M$_\odot$. Gonzalez et al. (2009) studied 417 red giants and supergiants in the Galactic bulge; only 13 stars (i.e. 3 per cent) showed the detectable Li~I 6708 line. Five of these 13 stars have $\log\epsilon$(Li) = 2.0-2.8. Their masses derived by us are $M$ = 2.8-5.2 M$_\odot$. 

% Table-4 
\begin{table}
  \caption{Upper limits for masses of Li-rich stars ($\log \epsilon$(Li) $\ge$~2.0)}
  \begin{minipage}{80mm}
    \begin{tabular}{cll}
    \hline
      Number & $M/M_{\odot}$ & Sources\\
      of stars&&\\
      \hline
      6 &$\leq 3.4\mathrm{^a}$&Brown et al. (1989)\\
      17&$\leq 4.7$           &Charbonnel \& Balachandran (2000)\\
      3 &$\leq 4.1\mathrm{^a}$&Reddy \& Lambert (2005);\\
        &                     &Kumar \& Reddy (2009)\\
      5 &$\leq 3.1\mathrm{^a}$&L\`{e}bre et al. (2006) \\
      5 &$\leq 5.2\mathrm{^a}$&Gonzalez et al. (2009)\\
      10&$\leq 5.3\mathrm{^a}$&present work\\
      \hline
    \end{tabular}
  \end{minipage}
$\mathrm{^a}$Masses of these stars are determined by us using Claret's (2004) 
evolutionary tracks and parameters $T_{\rm eff}$ and $\log g$ from the papers cited

\end{table}

      We present in Table~4 the upper limits for masses of Li-rich stars according to the cited papers including the present work (6 works in all). One may see that the extreme mass is $M$ = 5.3 M$_\odot$; the total $M$ range spreads from 1.0 to 5.3 M$_\odot$. Therefore, according to Table~4, all known Li-rich stars have masses $M <$~6~M$_\odot$. In other words, the Li-rich giants and supergiants are the relatively low-mass objects.

% #9
\section{COMPARISON WITH THEORETICAL PREDICTIONS}
% #9.1
\subsection{Recent computations of 2-15 M$_\odot$ stellar models}

      It was mentioned above that lithium is a much more sensitive indicator of rotational mixing than elements of participating in the CNO-cycles. For instance, according to computations by Heger \& Langer (2000) for a star with mass $M$~=~12~M$_\odot$ and initial rotational velocity $v_0$ = 200 km~s$^{-1}$, the surface nitrogen abundance by termination  of the MS phase increases by 0.4 dex, whereas the lithium abundance decreases by 8.0 dex. Recently, Frischknecht et al. (2010) performed computations of 9, 12 and 15 M$_\odot$ stellar models with rotational mixing and confirmed that the surface Li abundance decreases strongly during the MS phase. Recently Ekstr\"{o}m et al. (2012) enlarged significantly the  mass range of rotational stellar models by computing  models with $M$ from 0.8 to 120 M$_\odot$. Unfortunately, no predictions for the Li surface abundance are presented in this first of an expected series of papers.

      At our request, Urs Frischknecht extended computations to models with lower masses down to $M$ = 2 M$_\odot$. An initial lithium abundance $\log\epsilon$($^{7}$Li) = 3.25 was adopted in the  model computations. It is necessary to note that the first dredge-up (FD) phase is computed only partially in his models for masses $M$ = 7-15 M$_\odot$; the red-blue loops, which are pronounced for such masses, were not considered. We may compare our results with these new theoretical predictions (Frischknecht et al. 2010; Frischknecht, private communication) for rotating and non-rotating stellar models with masses $M$ from 2 to 15 M$_\odot$. 

% Fig-8
\begin{figure}
\begin{center}
\includegraphics[width=83mm]{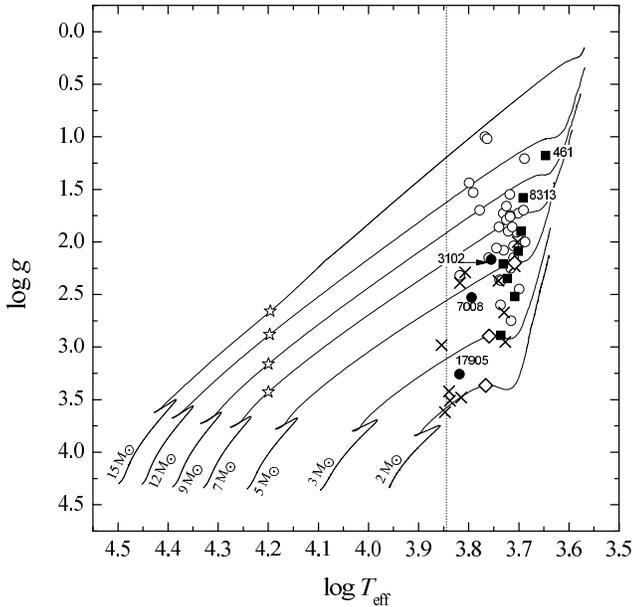}
\end{center}
\caption[]{Frischknecht's evolutionary tracks for non-rotating models and positions of programme stars on the $\log T_{\rm eff} - \log g$ plane. Symbols of the stars are the same as in Fig.7. Open rhombi on the 2, 3 and 5 M$_\odot$ tracks mark the beginning of the surface Li depletion in the FD phase. Open stars on the 7-15 M$_\odot$ tracks mark the post-MS moments when a great drop in $\log\epsilon$(Li) occurs. Dotted vertical line at $T_{\rm eff}$ = 7000~K separates the region of F and G supergiants from the region of A supergiants.}
\end{figure}

      A set of Frischknecht's evolutionary tracks for non-rotating 2-15 M$_\odot$ models is presented in Fig.8. These tracks are shown from the ZAMS to helium ignition and, hence, the red-blue loops for helium burning are absent (compare tracks in Figs 6 and 8). The dotted vertical line in Fig.8 corresponds to the effective temperature of 7000~K, the approximate boundary between A- and F-type supergiants. We found that stellar masses $M$ inferred from these new tracks are close to the masses derived from Claret's (also non-rotating) tracks (Fig.6). In particular, the difference in $M$ is imperceptible for stars with $M \sim$~2~M$_\odot$; the $M$ increment is larger for stars with $M \approx$ 5-11 M$_\odot$, namely it is about 8-11 per cent or comparable with the mean error of 8 per cent in the masses $M$ in Table~1.

      Insignificant differences in $M$ were obtained, when comparing masses derived from rotating and non-rotating models. We found that Frischknecht's tracks computed for the initial rotational velocities $v_0$ = 0 km s$^{-1}$ and 200 km s$^{-1}$ are rather close. Substitution of $v_0$ = 0 km s$^{-1}$ for 200 km s$^{-1}$ leads to a decrease of $M$ by 0.1 M$_\odot$ for a star with $M$~=~7.9~M$_\odot$ and by 0.3 M$_\odot$ for a star with $M$~=~~11.0~M$_\odot$. Thus, changes in $M$ are smaller than errors in the $M$ determinations. Therefore, use of models with the rapid initial rotation cannot lead to a marked decrease of the masses $M$.
 
      Stellar models with the initial rotational velocity $v_0$ = 0 km~s$^{-1}$ do not show any changes in the surface Li abundance during the MS phase. The changes appear later; according to Frischknecht's computations for non-rotating models, (i) if $M \le$~6~M$_\odot$ the surface Li abundance begins to decrease when a star enters the FD phase; onset of this decrement is marked in Fig.8 by open rhombi on the 2, 3, and 5 M$_\odot$ tracks (changes in the C and N surface abundances begin later, on the ascending branch of the tracks); (ii) if $M >$~6~M$_\odot$  strong Li depletion appears much earlier; these moments are marked by open stars on the 7, 9, 12 and 15 M$_\odot$ tracks. Therefore, both our finding (Fig.7) and recent theoretical predictions show that the stars studied can be divided into two groups depending on the mass $M$, namely the stars with $M \le$~6~M$_\odot$ and the stars with $M >$~6~M$_\odot$. Next, we discuss a comparison with the theoretical predictions for these two groups separately.

% #9.2
\subsection{Stars with masses $M \le$  6~M$_\odot$}

All Li-rich stars are found to be in this $M$ region. It is interesting that the Li-rich giants HD 17905 and HR 7008 have the Li abundances that are virtually equal to the initial abundance $\log\epsilon$(Li) = 3.2$\pm$0.1. A simple explanation of this fact arises, if one supposes that during the MS phase these stars had very slow rotation; they have not reached the FD phase (Figs 6 and 8) and, therefore, have conserved the initial Li abundance.

      The evolutionary status of the three Li-rich giants from Table~1 is clear from Fig.8 (they are marked there). The stars HD~17905 and HR~7008 are observed prior to reaching the FD phase, whereas the star HR~3102 is closer to it. Can the Li-rich giants HD~17905 and HR~7008 conserve their initial lithium abundance $\log\epsilon$(Li) $\approx$ 3.2 (the simplest explanation of their high Li abundances)? This would be possible, if their initial rotational velocities $v_0$ were close to zero. However, we found that these stars have currently rather high rotational velocities, namely $v \sin i$ = 53 and 58 km~s$^{-1}$, respectively (for the third Li-rich star, HR~3102, we found $v \sin i$ = 12 km~s$^{-1}$). It is known that when a B-type star terminates the MS phase and passes into the AFG supergiant phase, its rotational velocity decreases significantly. So, at the beginning of the MS phase, when the F-type giants HD~17905 and HR~7008 were B-type dwarfs, their rotational velocities should have been markedly greater. One may expect that the stars had then $v_0 \ge$ 100 km~s$^{-1}$ (see, e.g., Abt, Levato \& Grosso 2002). But stars with $v_0 \ge$ 100 km~s$^{-1}$ are predicted to decrease significantly their surface Li abundances by the end of the MS phase (see below). Then one must ask: Why has Li retained its initial value for HD~17905 and HR~7008? Are the predictions for rotationally-induced mixing in error? Are we observing freshly-made Li? Accurate determinations of C, N, and O abundances would be valuable here. 

      The third Li-rich star, the giant HR~3102 with $\log\epsilon$(Li) = 2.3, is of special interest because it is the only star in Table~1 for which the nitrogen abundance was obtained by Lyubimkov et al. (2011). Specifically, a nitrogen excess [N/Fe] = 0.41 was found for it. (Other stars in the cited  work were generally hotter than stars in the present work, i.e., Li could not be measured in these stars for which N
was measured.) According to Lyubimkov et al. (2011), one may consider two possible explanations for this N excess. First, the star with an initial high rotational velocity $v_0 \sim$ 200 km~s$^{-1}$ is now a post-MS object but has not yet reached the FD phase. However, Frischknecht's computations show that in this case the star should have very low, undetectable Li abundance. Second, the N excess would appear if the star had slow rotation on the ZAMS (e.g., $v_0 \sim$ 50 km/s; currently its observed velocity is $v \sin i$ = 12 km/s) and now is close to the FD phase termination. But in this case theory again predicts a Li abundance that is much less than the observed value $\log\epsilon$(Li) = 2.3. Moreover, as seen from Fig.6, this 5.1 M$_\odot$ star has not reached the FD phase. Thus, the observed data on the N and Li abundances cannot be explained by current theory. Therefore, it appears that this Li-rich bright giant HR~3102 may have freshly-produced lithium at the surface.

      In Table~3 there are three more Li-rich bright giants, namely HR 1135, HR 1644 and HR 6707, with the lithium abundances $\log\epsilon$(Li) = 2.2, 2.4 and 2.0, respectively. Their positions in Fig.8 show that they have not yet reached the FD phase. For one of these stars, namely HR 1135 (F3 II), the projected rotational velocity is known: $v \sin i$ = 49 km/s (Table~3). One may expect that the initial rotational velocity of the star $v_0$ was significantly greater; so we have in this case the same situation as for the above Li-rich giants HD 17905 and HR 7008. Since stellar models with rapid rotation predict a strong Li depletion by the end of the MS phase, one may suppose for this star, like HD 17905 and HR 7008, a presence of freshly-produced lithium.

      Four Li-rich giants from Table~3 (lower part) are interesting, because  they have similar low masses $M \approx$~2~M$_\odot$ and, simultaneously, the similar Li abundances $\log\epsilon$(Li) = 2.3-2.9. These giants are form a compact group in Figs 6, 7 and 8. One sees from Table~3 that their projected rotational velocities are $v \sin i$ = 27-66 km~s$^{-1}$, i.e. $\sim$ 50 km~s$^{-1}$ on average. They all are post-MS objects but have not reached the FD-phase (Fig.8). Frischknecht's computations for a 2 M$_\odot$ model with the initial rotational velocity $v_0 \sim$ 50 km~s$^{-1}$ predict the Li abundance $\log\epsilon$(Li) $\sim$ 2.6 just after the MS phase. This value is in excellent agreement with the observed Li abundances for these giants. Therefore, their enhanced Li abundance is appears explainable. However, if, as seems likely, rotational velocities on ZAMS were significantly greater than current values, one may have to suppose the presence of freshly-produced lithium. This supposition would be confirmed, if, as for HR 3102, a nitrogen excess was found for these giants.

      By the FD termination, the models with $v_0$ = 0 km~s$^{-1}$ predict the surface Li abundance to fall in a narrow range from $\log\epsilon$(Li) = 1.46 at $M$ = 2 M$_\odot$ to $\log\epsilon$(Li) = 1.37 at $M$ = 6~M$_\odot$. This relation between $\log\epsilon$(Li) and $M$ is shown by the solid thick line in Fig.7.  The mean predicted abundance is $\log\epsilon$(Li) = 1.4 which is equal to the mean observed $\log\epsilon$(Li) value for eleven stars with $\log\epsilon$(Li) = 1.1 to 1.8 (six stars from Table~1, filled squares in Fig.7, and five stars from Table~3, crosses in Fig.7). Thus, we arrive at the conclusion that these eleven stars can be explained as post-FD objects (or objects well into the FD phase) that had small rotational velocities on the ZAMS. Positions of these stars in Fig.8 confirm that they are in the FD phase (some may be on a red-blue loop). Their current rotational velocities are very low ($v \sin i <$ 5 km/s, see Table~3) Thus, for 11 stars of our sample the initial rotational velocity is expected to be very small ($v_0 \sim$ 0 km/s). This fact leads to the conclusion that a substantial fraction of B dwarfs, progenitors of F and G supergiants and bright giants, have low rotational velocities $v_0$ that are close to zero.

       In the case of 2-6~M$_\odot$ models with rotation,  substantial lithium depletion in the atmospheres occurs already in the MS phase and continues during the FD phase. We show in Fig.7 by the dashed line the predicted post-FD surface abundances for 2-6~M$_\odot$ models with the initial rotational velocity of $v_0$ = 50 km~s$^{-1}$. One sees that the Li abundance decreases from $\log\epsilon$(Li) = 0.92 at $M$ = 2 M$_\odot$ to $\log\epsilon$(Li) = 0.39 at $M$ = 6~M$_\odot$. It is hardly possible to detect such low abundances. When $v_0$ = 100 km~s$^{-1}$, the theory predicts severe Li depletion already at MS termination, specifically from $\log\epsilon$(Li) = 0.80 at $M$ = 2 M$_\odot$ to $\log\epsilon$(Li) = -0.85 at $M$ = 6~M$_\odot$. During the FD phase a further strengthening of Li depletion occurs. Therefore, most of our 2-6~M$_\odot$ stars with Li upper limits (open circles in Figs 7 and 8) can be explained as post-FD objects with $v_0 \approx$ 50 km~s$^{-1}$ or post-MS and post-FD objects with $v_0 \ge$ 100 km~s$^{-1}$. 

      One sees that two open circles (Li abundance upper limits) in Fig.7 lie higher than the post-FD predictions (solid line at $\log\epsilon$(Li) $\approx$ 1.4); they correspond to the F stars HR~7542 and HR~7834, both with a mass $M$ = 5.3 M$_\odot$. Their high upper Li limits (higher than Li limits for other stars of this group) can be explained by difficulties in detecting the Li~I 6708 line. In particular, the spectrum of HR~7542 is markedly smoothed due to the rotational velocity $v \sin i$ = 25 km~s$^{-1}$. A weak asymmetric blend observed at 6708~\AA\ may be partially due to the Li~I line, but the appreciable contribution of nearby Fe I and Si I lines does not allow us to estimate reliably the $\log\epsilon$(Li) value. As far as HR~7834 is concerned, its high Li upper limit is related to its    rather high effective temperature $T_{\rm eff}$ = 6570~K (see Fig.5). Positions of both stars in Fig.8 show that they have not reached the FD phase. The bright giant HR~7834 may  be a post-MS object with, e.g., $v_0$ = 100 km~s$^{-1}$ and with the predicted surface abundance $\log\epsilon$(Li) = -0.7 (the post-MS theoretical value interpolated to $v_0$ = 100 km~s$^{-1}$ for the mass $M$ = 5.3 M$_\odot$). The supergiant HR~7542 has currently the relatively high rotational velocity $v \sin i$ = 25 km~s$^{-1}$; its initial velocity $v_0$ was significantly greater. If its Li upper limit 1.5 is close to the actual Li abundance, one may suppose for this star the same scenario as for the HR~1135 (see above), i.e. an appearance of freshly-produced lithium.

% #9.3
\subsection{Stars with masses $M >$~6~M$_\odot$}

In contrast to the preceding group, the derived Li abundances for stars with $M >$~6~M$_\odot$ display less variety. One may see from Fig.7 that Li-rich stars are absent in the $M$ region from 6 to 15 M$_\odot$. As mentioned above, data of other authors confirm that there are no Li-rich stars with $M >$~6~M$_\odot$. Next, only  two stars, namely HR~461 and HR~8313 (marked in Fig.7), have detectable $\log\epsilon$(Li) values. In all other stars of this group lithium is not seen. 

      As far as the theoretical predictions on Li depletion are concerned, a dramatic difference takes place for the non-rotating ($v_0$ = 0 km~s$^{-1}$) models with masses $M$ = 7-15 M$_\odot$ in comparison with the 2-6~M$_\odot$ models. Although the non-rotating 7-15 M$_\odot$ models, like similar 2-6~M$_\odot$ models, do not show any changes in the surface $\log\epsilon$(Li) value during the MS phase, a deep and precipitate drop in $\log\epsilon$(Li) by 4-6 dex occurs shortly after the MS termination (its value is dependent on $M$). We mark in Fig.8 these very short moments by open stars on the 7-15 M$_\odot$ tracks. (According to the model computations, the drop is absent at $M$ = 6 M$_\odot$, but exists at $M$ = 7 M$_\odot$).  These abrupt changes in $\log\epsilon$(Li) occupy a very narrow range of effective temperatures in Fig.8, namely from $T_{\rm eff}$ = 15820~K at $M$ = 7 M$_\odot$ to $T_{\rm eff}$ = 15700~K at $M$ = 15 M$_\odot$. Therefore, a great drop in $\log\epsilon$(Li) occurs, when a star is observed as a B-type giant, i.e. when it is rather far from the F and G supergiant region where $T_{\rm eff} \le$ 7000~K. Moving along the track, such a star comes to the F and G supergiant region with a very low, undetectable Li abundance. One may suppose that the abrupt drop of the surface Li abundance arises from short-lived mixing (unrelated to rotation) of the outer layers, and, therefore, leads to the lithium destruction but does not affect other chemical elements. 

      The 7-15 M$_\odot$ models with rotation predict significant Li depletion already at the MS phase termination. For instance, for the models with $v_0 \approx$ 100 km~s$^{-1}$ the surface abundance $\log\epsilon$(Li) at the end  of the MS varies from -0.9 for $M$ = 7 M$_\odot$ to -4.3 for $M$ = 15 M$_\odot$; for the models with $v_0 \approx$ 200 km~s$^{-1}$ this value varies from -5.7 to -6.8, respectively. Further lithium deficiency occurs when a star is evolving from a B giant to an F and G supergiant and, next, when it enters into the FD phase.
      Therefore, from the viewpoint of Frischknecht's computations, F and G supergiants and bright giants with  detectable Li abundance should not be observed in the 7-15 M$_\odot$ range.  However, there are two stars in the 7-15 M$_\odot$ range, namely HR~461 and HR~8313 (marked in Fig.7), with a definite $\log\epsilon$(Li) value. It is necessary to consider these stars in more detail. 

      The Li~I 6707.8 line is clearly seen in spectra of both stars (see, e.g., Fig.1 for HR~8313). The non-LTE lithium abundances of HR 461 and HR 8313 (Table~1) agree very well with the mean post-FD predicted abundance $\log\epsilon$(Li) = 1.4 for non-rotating models with $M \leq$~6~M$_\odot$ (thick solid line in Fig.7), as well as with the mean observed $\log\epsilon$(Li) value for the 11 stars in the less massive group discussed above (six filled squares and five crosses). Our $\log\epsilon$(Li) values for the stars HR~461 and HR~8313 show good agreement with data of other authors. In particular, Luck \& Wepfer (1995) derived the LTE abundance $\log\epsilon$(Li) = 0.92 for HR~461, which agrees very well with our LTE value 0.89 for this star (Table~1). Luck (1977) obtained the non-LTE abundance $\log\epsilon$(Li) = 1.49 for HR~8313 that is in excellent accordance with our value 1.53. 

      Masses of the stars HR~461 and HR~8313 were found to be $M$ = 9.5$\pm$1.2 and 7.1$\pm$0.4 M$_\odot$, respectively (Paper I). When determining the masses, we took our $T_{\rm eff}$ and $\log g$ and used Claret's (2004) evolutionary tracks, which have been computed with no rotational mixing in the MS phase. (Note that errors in the $M$ values in Paper I were estimated from uncertainties in $T_{\rm eff}$ and $\log g$). When masses are derived from the new Frischknecht's tracks presented in Fig.8, we obtain the slightly increased $M$ values, namely 10.6~M$_\odot$ for HR~461 and 7.9 M$_\odot$ for HR~8313. 

      Although it is difficult to suppose that the derived masses $M$ for the stars HR~461 and HR~8313 contain gross errors, the theoretical predictions are such that cool giants and supergiants with $M >$~6~M$_\odot$ cannot display a detectable Li abundance (this is possible only for stars with $M \le$~6~M$_\odot$). Therefore, one may suggest two hypotheses as follows: 

   (i) Lithium observed in the atmospheres of the stars HR~461 and HR~8313 was produced recently, e.g., during the FD phase. One sees from Fig.8 that both stars are in the FD phase or, maybe, close to its termination (if they are on the red-blue loops). Note in this connection that the G-type supergiant HR~8313 shows the carbon isotope ratio $^{12}$C/$^{13}$C = 34 (see below), which is typical for stars coming to the FD's end. 

   (ii) The prediction of the non-rotating models with $M >$~6~M$_\odot$ that a great drop in $\log\epsilon$(Li) occurs shortly after the MS phase is incorrect, but rather stars with $M$ = 7-10 M$_\odot$ and $v_0 \sim$ 0 km~s$^{-1}$, like similar 2-6~M$_\odot$ stars, have the post-FD lithium abundance about $\log\epsilon$(Li) = 1.4. In this case the stars HR~461 and HR~8313 form in Fig.7 a common sequence with the six stars of lower masses $M <$~6~M$_\odot$ (filled squares). One may suppose that all these stars passed through the FD phase and, therefore, received the Li depletion that is typical for the post-FD phase. 

     The predicted Li abundance is rather sensitive to details of stellar model computations. Comparison of Frischknecht's results with earlier computations of Heger \& Langer (2000) for models of rotating stars with masses $M$ = 12 and 15 M$_\odot$ confirms this supposition. In fact, when comparing data from these two sources obtained for initial rotational velocities $v_0$ = 100 and 200 km~s$^{-1}$, we found that Frischknecht's models predict much larger reductions in $\log\epsilon$(Li) at the MS termination than Heger \& Langer's ones (the difference is about 2-4 dex). Note that a significant difference is also found by Frischknecht et al. (2010) for boron depletion. Frischknecht et al. (2010) noted that uncertainties in the theory are still quite large. Therefore, further  theoretical efforts are necessary in order to explain all data on the lithium depletion in post-MS and post-FD stars and, in particular, to confirm (or not) the predicted great drop in $\log\epsilon$(Li) shortly after the MS phase in the non-rotating 7-15 M$_\odot$ models. 

      Our analysis, as well as theoretical predictions shows that the surface Li abundances in F and G supergiants and giants depend strongly on their initial rotational velocities $v_0$. What actual $v_0$ values can be expected for such stars? In particular, can supergiants with the very low values $v_0 \sim$ 0 km~s$^{-1}$ exist? 

      Progenitors of the 3-15 M$_\odot$ supergiants are B-type MS stars, whereas the 2 M$_\odot$ supergiants' progenitors are A-type MS stars. Abt, Levato \& Grosso's (2002) analysis of observed rotational velocities of B-type MS stars led to interesting conclusions on the $v_0$ values. According to their results, there is a rather large number of B-type MS stars with low initial velocities $v_0 \sim$ 50 km~s$^{-1}$. In particular, for B0-B2 subtypes the fraction of stars with $v_0$ = 50 km~s$^{-1}$ is greater than the fraction with $v_0$ = 100 or 200 km~s$^{-1}$, whereas for B6-B7 subtypes the number of stars with $v_0$ = 50 km~s$^{-1}$ is comparable with the number of stars with $v_0$ = 200 km~s$^{-1}$. Therefore, the existence of F and G supergiants and giants both with the relatively small initial values $v_0 \sim$ 50 km~s$^{-1}$ and with the relatively large values $v_0 \ge$ 100 km~s$^{-1}$ is possible. Moreover, since for 11 stars with $\log\epsilon$(Li) $\sim$ 1.4 and $M <$~6~M$_\odot$ (6 filled squares and 5 crosses in Fig.7) the values $v_0 \sim$ 0 km~s$^{-1}$ are expected (see above), the fraction of stars with very low initial velocities $v_0$ is substantial, too. 

      Recently Huang, Gies and McSwain (2010) determined and analysed rotational velocities for a large sample ($\sim$1000) of B-type MS stars. They considered evolution of rotation of the stars during the MS phase (from ZAMS to TAMS). It was concluded that about 6 per cent of newborn B stars are very slow rotators, i.e. they have the ratio $V_{eq}/V_{crit} <$ 0.1, where $V_{eq}$ and $V_{crit}$ are equatorial and critical rotational velocities, respectively. However, the distribution of the stars with $V_{eq}/V_{crit}$ seems to depend strongly on stellar mass $M$, so as the mass $M$ increases, more and more stars are born as slow rotators. The fraction of slow rotators is relatively small for newborn B stars with $M$ = 2-4 M$_\odot$, but is large for the stars with $M >$ 8 M$_\odot$. In particular, about 84 per cent of the stars with $M >$ 8 M$_\odot$ show the initial ratio $V_{eq}/V_{crit} <$ 0.5 and a significant part of them have $V_{eq}/V_{crit} <$ 0.2 (see Fig.7 in the work cited). Therefore, Huang, Gies and McSwain's analysis confirms that progenitors of F and G supergiants and bright giants could have the rotational velocity distribution needed to account for the distribution of Li abundances with mass.

% #10
\section{CONCLUDING REMARKS}

Computations of stellar evolutionary models show that lithium is a very sensitive indicator of stellar evolution. During evolution of a star from the main sequence (MS) to the F and G supergiant phase, lithium may be destroyed by, for example, the rotationally-induced mixing in the MS phase and strongly diluted by development of the supergiant's convective envelope (the first dredge-up, FD). The strong dilution of the surface Li abundance in rotational models by the MS phase termination and then during the FD phase is predicted, e.g., by computations of Heger and Langer (2000) and Frischknecht et al. (2010).

      Observed lithium abundances in F and G supergiants and giants support the idea that there is a boundary at $M \approx$~6~M$_\odot$ that divides these stars  into two groups. In particular, all known Li-rich giants and supergiants seem to be stars with masses $M <$~6~M$_\odot$; there are no Li-rich stars with $M >$~6~M$_\odot$. This conclusion follows not only from our work but also from a scrutiny of published data. A division of the stars into two groups with $M \le$~6~M$_\odot$ and $M >$~6~M$_\odot$ is suggested as well by the recent Frischknecht's computations (priv. comm.) of stellar models with masses from 2 to 15 M$_\odot$. 

      Most of the stars studied do not show lithium in their spectra. Low lithium abundances in their atmospheres are explained by the strong Li depletion during the star's evolution; a few possibilities can be suggested: 

      (i) In stars with masses $M >$~6~M$_\odot$ and very slow rotation ($v_0 \sim$ 0 km~s$^{-1}$) a great drop in $\log\epsilon$(Li) is predicted to occur shortly after the MS phase, such that stars come to the F and G supergiant phase with a strong Li deficiency. In rotating stars with $M >$~6~M$_\odot$ the great Li deficiency appears already by the MS phase termination, according to Frischknecht; the deficiency increases when a star passes through the FD phase. So, from the viewpoint of these theoretical predictions, F and G supergiants and bright giants with $M >$~6~M$_\odot$ cannot display a detectable Li abundance. 

      (ii) In stars having masses $M \le$~6~M$_\odot$ and initial rotational velocities $v_0 \ge$ 100 km~s$^{-1}$ the significant Li depletion is predicted to appear by the end of the MS phase. When these stars become F and G supergiants and giants, Li is predicted to be undetectable. The Li abundance becomes too low as well for stars with $v_0 \sim$ 50 km~s$^{-1}$, when they terminate the FD phase. 

      There are altogether 23 stars (of 55 studied) with detectable Li abundances, namely 11 stars in our basic list (Table~1) and 12 stars in the additional list (Table~3). For eleven of these stars with masses $M <$~6~M$_\odot$ we found $\log\epsilon$(Li) in a rather narrow range from 1.1 to 1.8 (the average is 1.4). These Li abundances agree very well with the mean predicted value $\log\epsilon$(Li) = 1.4 for the post-FD non-rotating 2-6 M$_\odot$ models. So, these 11 stars can be explained as the post-FD 2-6 M$_\odot$ objects with the very slow initial rotation ($v_0 \sim$ 0 km/s).

      We studied 10 Li-rich stars (3 in Table~1 and 7 in Table~3); their Li abundances are $\log\epsilon$(Li) = 2.0-3.2. We found that all Li-rich giants and supergiants with $\log\epsilon$(Li) $\geq$ 2.0 have masses $M <$~6~M$_\odot$; this conclusion follows not only from our work but also from a scrutiny of published data.

      Two Li-rich giants in Table~1 caught prior to the FD, namely HD~17905 and HR~7008, have a Li abundance that is close to their initial/interstellar value $\log\epsilon$(Li) = 3.2. Theory insists that to have conserved their initial Li abundance, their initial rotational velocities $v_0$ must have been about zero. However, their observed rotational velocities as bright giants are $v \sin i$ = 50-60 km~s$^{-1}$ and, therefore, the $v_0$ values at the MS beginning are expected to be much greater. Rotating stellar models with $v_0 \ge$ 100 km~s$^{-1}$ predict significant Li deficiency after the MS phase; the observed high abundances $\log\epsilon$(Li) for HD~17905 and HR~7008 contradict the theory. The same situation occurs for the bright giant HR~1135 (F3 II) from Table~3 with $\log\epsilon$(Li) = 2.2 and with the otational velocity $v \sin i$ = 49 km~s$^{-1}$. Therefore, we are forced to assume for these three stars that either the theoretical Li deficiency is greatly overestimated or freshly-produced lithium is observed in their atmospheres. Lithium production is likely needed as well for one more Li-rich star in our sample, namely HR~3102 (Table~1), because the theory, even for non-rotating stars, cannot explain simultaneously its high lithium abundance and substantial nitrogen excess. 

      There is also in Table~3 a compact group of four Li-rich giants (luminosity class III) with  masses $M \approx$ 2 M$_\odot$ and Li abundances $\log\epsilon$(Li)~=~2.3-2.9. The observed $\log\epsilon$(Li) values for these four post-MS stars would be in good agreement with the theory, if their current rotational velocities as giants $v \sin i \sim$ 50 km~s$^{-1}$ do not markedly differ from the initial values $v_0$ on ZAMS. However, if the $v_0$ values were significantly greater than current velocities, one may suppose again the presence of freshly-produced lithium. 

      The absolute majority of programme stars with $M >$~6~M$_\odot$ did not show detectable lithium that is in good agreement with Frischknecht's predictions. Only for two stars with $M >$~6~M$_\odot$, namely HR~461 and HR~8313, did we find  definite $\log\epsilon$(Li) values. Present theory cannot explain the detectable Li abundance in their atmospheres. Two possible solutions of the problem may be suggested. First, the predictions for non-rotating 7-15 M$_\odot$ models of a great drop in surface Li abundance shortly after the MS phase are invalid. In this case, the stars HR~461 and HR~8313, like eleven 2-6~M$_\odot$ stars with $\log\epsilon$(Li) $\approx$ 1.4, can be post-FD objects with $v_0 \sim$ 0 km~s$^{-1}$. Second, the theory is valid, but an episode of Li production has been overlooked. 

      Evolutionary status of the stars studied would be clearer if additional information was available on abundances of elements and isotopes predicted to be affected by mixing with the interior.  The carbon isotope ratio $^{12}$C/$^{13}$C is a valuable indicator but requires a C-containing molecule to contribute to the stellar spectrum. Unfortunately, we could find published $^{12}$C/$^{13}$C values for only two cool supergiants, namely HR~3045 ($\xi$ Pup) and HR~8313 (9~Peg). According to Tomkin, Luck and Lambert (1976), the $^{12}$C/$^{13}$C ratio is equal to 7 for HR~3045 and 34 for HR~8313 that is much less than the initial (solar) ratio $^{12}$C/$^{13}$C = 89. The low $^{12}$C/$^{13}$C values confirm that both the stars have been through the FD phase; both are rather massive ($M$ = 9.9 and 7.1 M$_\odot$, respectively), so they are likely to be on their red-blue loops. As follows from Frischknecht's computations, the extremely low ratio $^{12}$C/$^{13}$C = 7 for the post-FD 9-15 M$_\odot$ stars, like that for HR~3045, can be explained by evolution with an initial rotational velocity $v_0$ = 150 km~s$^{-1}$. 

      Observational insight into the changes in surface abundances of the progenitors of FGK supergiants and giants would be valuable. Deep mixing such as may alter the surface C, N, and O abundances is guaranteed to rid the surface of lithium but  lithium (as explained above) is not observable in the hot progenitors. Greater exploration of boron in such stars could be very helpful, and beryllium in 
stars of intermediate temperatures would be valuable too. Boron as B~II and B~III is 
detectable only in the ultraviolet and beryllium as Be~II only at 3130~\AA. 

      Summarising, we may conclude that current theoretical stellar models with and without rotation account fairly well for the observed Li abundances in F and G giants and supergiants. Nonetheless, there are cases where the theory may be invalid, namely (i) the Li-rich giants HD~17905, HR~7008, HR~3102 and HR~1135 ($M <$~6~M$_\odot$), whose high Li abundances accompanied with rather large rotational velocities (HD~17905, HR~7008 and HR~1135) or the substantial nitrogen excess (HR~3102) contradict theoretical predictions; (ii) the relatively high-mass supergiants HR~461 and HR~8313 ($M >$~6~M$_\odot$) with the detectable abundances $\log\epsilon$(Li) = 1.3-1.5. It is possible that recently synthesized lithium is observed in both cases (i) and (ii). 

      In all discussions of lithium synthesis in evolved stars, the same recipe involving nuclear reactions is invoked: ${^3}$He~+~$\alpha~\rightarrow~{^7}$Be~+~$\gamma$; ${^7}$Be~+~e${^-}~\rightarrow~{^7}$Li~+~$\nu_{\rm e}$. The $^{3}$He is that which was present at the birth of the star and escaped destruction by avoiding mixing with the interior. (In low-mass stars but not the progenitors of the FGK bright giants and supergiants. Additional and fresh $^{3}$He is synthesized by the first stages of the pp-chain). If the above recipe is to result in lithium production, it must operate in a convective environment such that $^{7}$Be and $^{7}$Li is convected to lower temperatures at which they may avoid destruction by protons. This pairing of synthesis and a convective environment is known as the Cameron-Fowler mechanism. In connection with red giants, three sites at which the Cameron-Fowler mechanism may operate have been previously proposed: red giants experiencing the 'bump' on the first ascent of the RGB or the He-core flash in low mass red giants (Charbonnel \& Balachandran 2000; Kumar et al. 2011), the ascent of the AGB following a red-blue loop (Charbonnel \& Balachandran 2000) and finally the hot bottom convective envelope of a luminous AGB star (Sackmann \& Boothroyd 2000). Here, the suggestion of lithium production following a red-blue loop might account for lithium in HR~461 and HR~8313. A new site and/or mechanism seems to be required to account for lithium production in HD~17905, HR~7008, HR~3102 and HR~1135 unless present theory overestimates destruction of Li by rotationally-induced mechanisms. Extra-mixing associated with a presence of magnetic fields has been suggested, for example (Guandalini et al. 2009).

\section*{Acknowledgments}

We thank Urs Frischknecht who placed at our disposal his results of evolutionary computations on the surface Li abundance for both rotating and non-rotating stellar models. We thank Ulrike Heiter for a helpful referee's report. DLL thanks the Robert A. Welch Foundation of Houston, Texas for support through grant F-634.

\label{lastpage}
\end{document}